\documentclass[11pt,onecolumn]{article}
\usepackage[bottom]{footmisc}
\usepackage[affil-it]{authblk}
\usepackage[margin=0.5in]{geometry}
\usepackage{xcolor,soul}
\usepackage{graphicx,caption,subcaption}
\usepackage[superscript,biblabel]{cite}
\usepackage{parskip,bm}
\usepackage{amsmath,amssymb,amsfonts,mathtools}
\usepackage{xspace}
\usepackage{hyperref}
\usepackage{flafter}

\DeclareMathOperator*{\argmax}{arg\,max}
\usepackage[noline,ruled]{algorithm2e}
\SetKwFor{eRepeat}{repeat}{}{}
\SetKwFor{myFor}{for}{do}{end}
\SetKwFor{myIf}{if}{then}{}
\SetKwFor{myElse}{else}{}{end}

\setstcolor{red}
\newcommand{\Ady}{$\mathit{IS}_{d}$\xspace}
\newcommand{\Aeta}{$\mathit{IS}_\eta$\xspace}
\newcommand{\Pdy}{$\pi_{d}$\xspace}
\newcommand{\Peta}{$\pi_\eta$\xspace}
\newcommand{\Sdy}{$\mathit{SS}_{d}$\xspace}
\newcommand{\Seta}{$\mathit{SS}_\eta$\xspace}

\newcommand{\MovieOne}{Supplementary Movie S1}
\newcommand{\MovieTwo}{Supplementary Movie S2}
\newcommand{\MovieThree}{Supplementary Movie S3}
\newcommand{\MovieFour}{Supplementary Movie S4}

\title{Efficient collective swimming by harnessing vortices\\ through deep reinforcement learning}

\renewcommand\footnotemark{\textsuperscript{$\dagger$}}

\author
{Siddhartha Verma \textsuperscript{$1\dagger$},  Guido Novati \textsuperscript{$1$}\footnote{\textsuperscript{$\dagger$}Authors contributed equally to this work}, Petros Koumoutsakos$^{1\ast}$\\
$\ $\\
\normalsize{$^{1}$Computational Science and Engineering Laboratory, Clausiusstrasse~33, ETH~Z\"{u}rich, CH-8092, Switzerland}\\
$\ $\\
\normalsize{$^\ast$Corresponding author e-mail: petros@ethz.ch}
}

\date{}

\begin{document} 

\maketitle 

\textbf{Classification}:  
\begin{enumerate}
\item Physical Sciences/Applied Mathematics
\item Biological Sciences/Biophysics and Computational Biology 
\end{enumerate}


\textbf{Short Title}: Efficient propulsion through deep reinforcement learning


\pagebreak

\begin{abstract}
Fish in schooling formations navigate complex flow-fields replete with mechanical energy in the vortex wakes of their companions. Their schooling behaviour has been associated with evolutionary advantages including collective energy savings. How fish harvest energy from their complex fluid environment and the underlying physical mechanisms governing energy-extraction during collective swimming, is still unknown. Here we show that fish can improve their sustained propulsive efficiency by actively following, and judiciously intercepting, vortices in the wake of other swimmers. This swimming strategy leads to collective energy-savings and is revealed through the first ever combination of deep reinforcement learning with high-fidelity flow simulations. 
We find that a `smart-swimmer' can adapt its position and body deformation to synchronise with the momentum of the oncoming vortices, improving its average swimming-efficiency at no cost to the leader. The results show that fish may harvest energy
deposited in vortices produced by their peers, and support the conjecture that swimming in formation is energetically advantageous. Moreover, this study demonstrates that deep reinforcement learning can produce navigation algorithms for complex flow-fields, with promising implications for energy savings in autonomous robotic swarms.
\end{abstract}

{\paragraph{Significance Statement.} \textit{Fish schooling is one of the most intriguing instances of collective behavior and complex decision making in nature, yet its underlying physical mechanisms remain largely unknown. We combine state of the art flow simulations with reinforcement learning, to answer the longstanding question of whether schooling fish may reduce energy-expenditure by adapting their swimming motion to the flow created by their companions. We demonstrate that a `smart' self-propelled swimmer can autonomously adapt its swimming behaviour to exploit energy deposited in the wake of other swimmers. The results support the thesis that fish may exploit unsteady flow-fields generated by collective locomotion to reap substantial energetic benefits and have promising implications for autonomous robotic swarms.}
}


There is a long-standing interest for understanding and exploiting the physical mechanisms employed by active swimmers in nature (nektons)~\cite{Schmidt1923,Pryor1966,Nekton1977,Triantafyllou2016}. 
Fish schooling in particular, one of the most striking patterns of collective  behaviour, has been the subject of intense investigations \cite{Breder1965,Weihs1973,Shaw1978,Pavlov2000,Burgerhout2013}. A key issue in understanding  fish schooling behaviour, and its potential for engineering applications \cite{Whittlesey2010}, is the clarification of the role of the flow environment.
Fish sense and navigate in complex flow-fields full of mechanical energy that is distributed across multiple scales by vortices generated by obstacles and other swimming organisms ~\cite{Chapman2011,Montgomery1997} .  There is  evidence that their swimming behaviour adapts to flow gradients (rheotaxis) and, in certain cases, it reflects energy-harvesting from such environments \cite{Lyon1904,Liao2003Science,Oteiza2017}. Hydrodynamic interactions have also been implicated in the fish schooling patterns that form when individual fish adapt their motion to that of their peers, while compensating for flow-induced displacements. Recent experimental studies have argued that fish may interact beneficially with each other ~\cite{Herskin1998,Killen2012,Burgerhout2013}, but in ways that challenge \cite{Ashraf2017} the  earlier proposed mechanisms ~\cite{Breder1965,Weihs1973} governing fish-schooling. However, the role of hydrodynamics in fish schooling is not embraced universally~\cite{Pitcher1986,Pavlov2000,Lopez2012} and there is limited quantitative information  regarding the physical mechanisms that would explain such energetic benefits.  Experimental~\cite{Herskin1998,Killen2012} and computational~\cite{Daghooghi2015} studies of collective swimming have been hampered by the presence of multiple deforming bodies and their interactions with the flow-field. Moreover, numerical simulations have demonstrated that a coherent swimming group cannot be sustained without exerting some form of control strategy on the swimmers~\cite{Gazzola2014,Maertens2017}. Here, we employ deep reinforcement learning (deep RL~\cite{Mnih2015}) to discover such strategies for two autonomous and self-propelled swimmers, and elucidate the physical mechanisms that enable efficient and sustained coordinated swimming.

\begin{figure}
	\centering
	\begin{subfigure}[b]{15cm}
		\centering
		\includegraphics[width=0.8\textwidth]{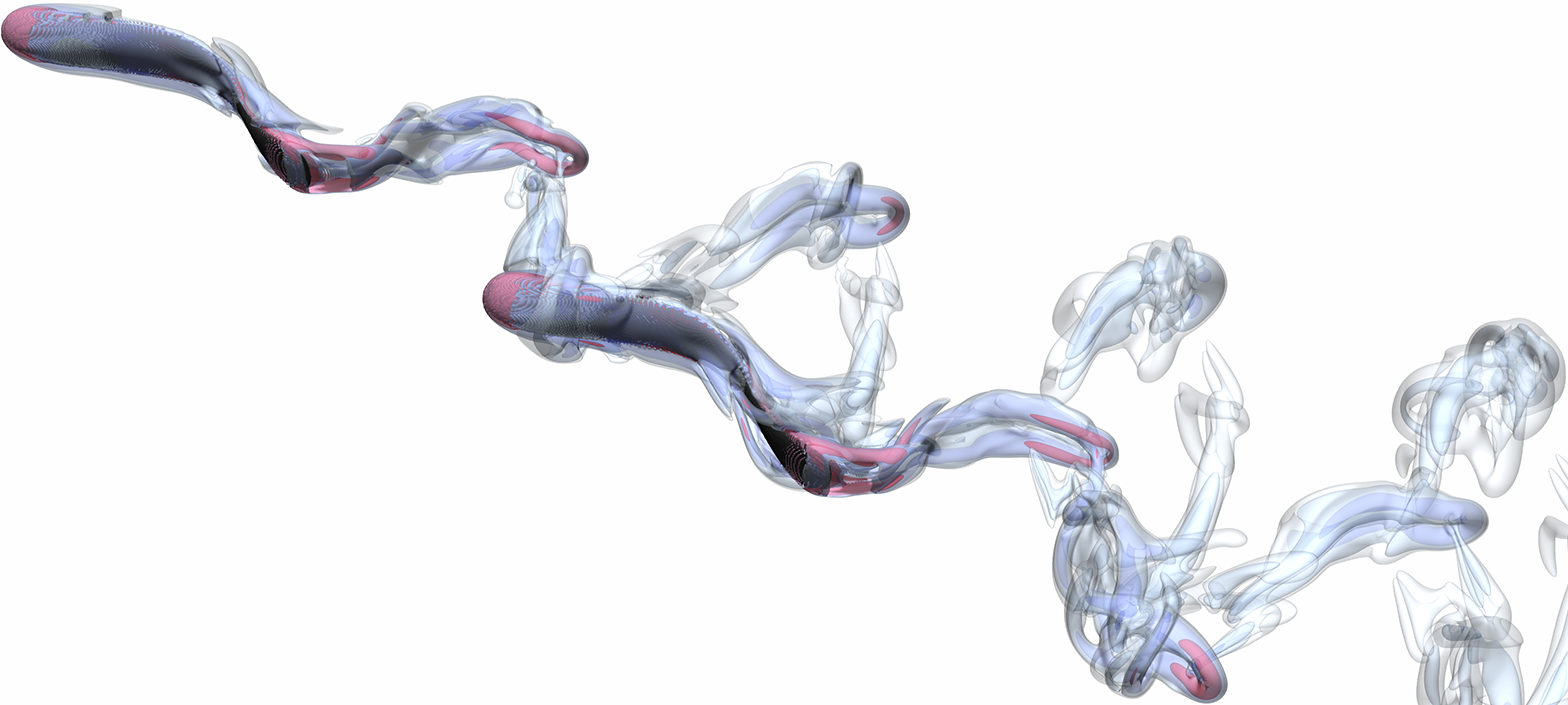}
		\subcaption{}
		\label{fig:3D2fish}
	\end{subfigure}
	\begin{subfigure}[b]{15cm}
		\centering
		\includegraphics[width=0.8\textwidth]{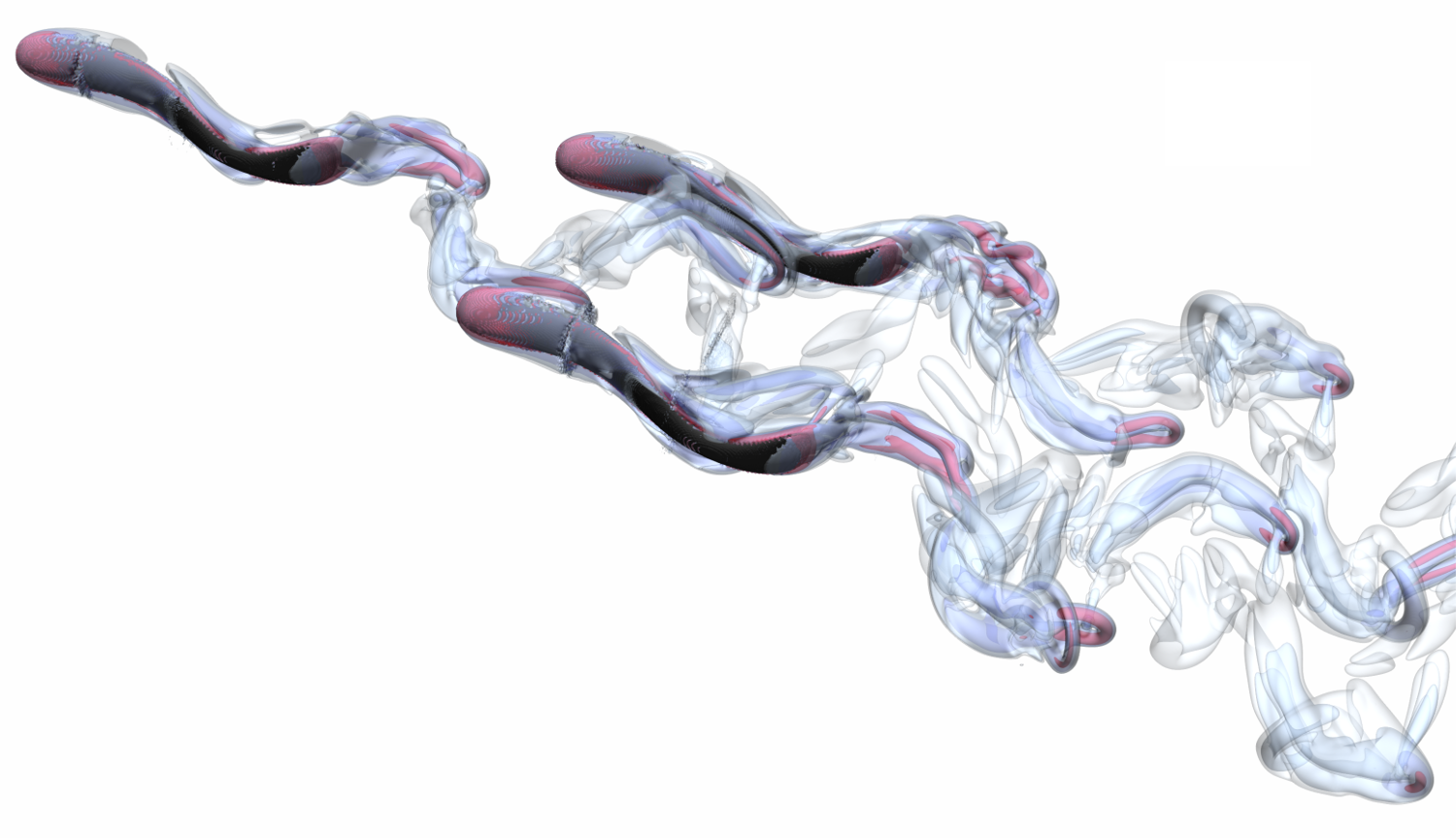}
		\subcaption{}
		\label{fig:3Dpair}
	\end{subfigure}
	\begin{subfigure}[b]{8.9cm}
		\centering
		\includegraphics[width=1.0\textwidth]{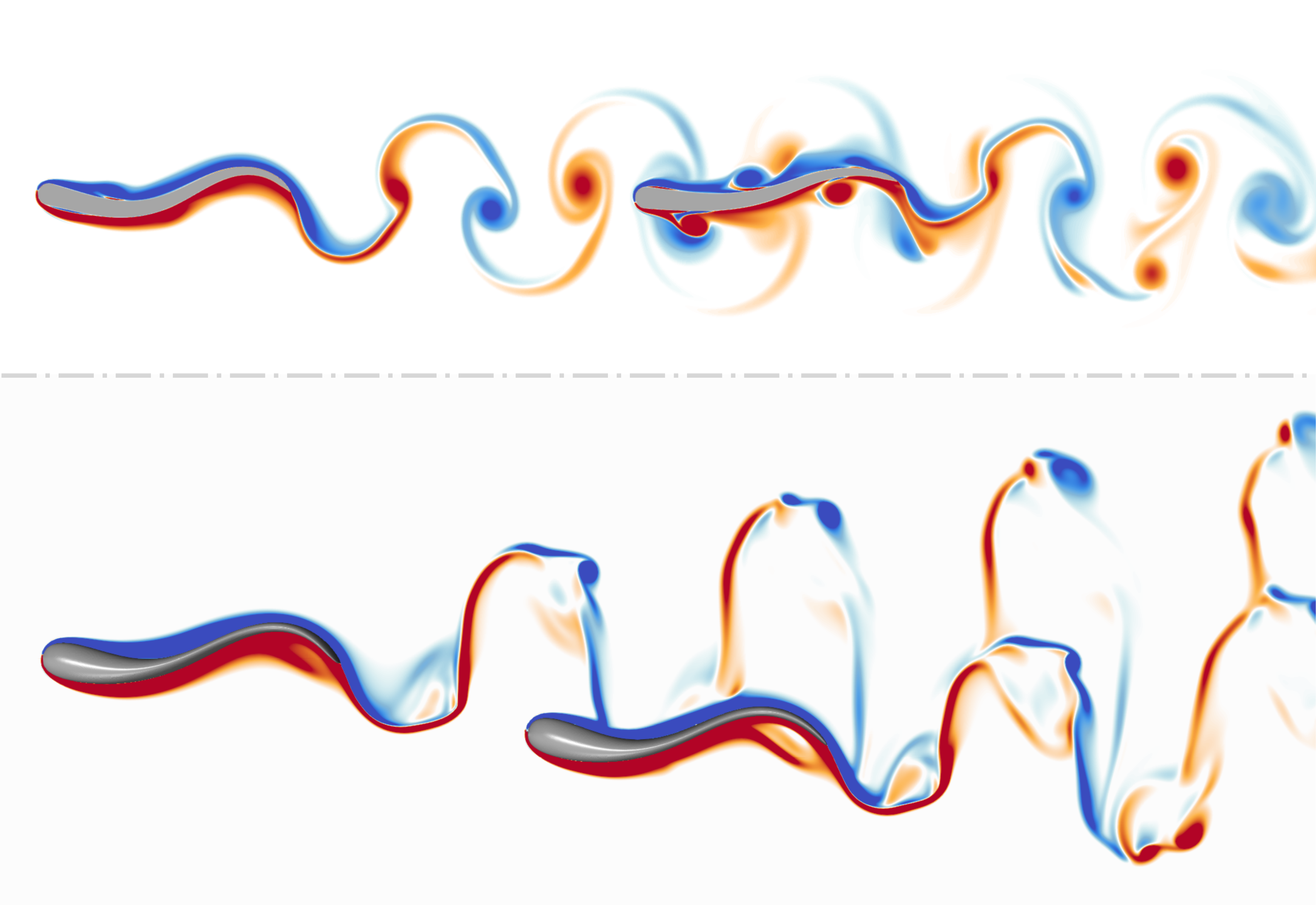}
		\subcaption{}
		\label{fig:goldenRatio}
	\end{subfigure}
	\begin{subfigure}[b]{8.9cm}
		\centering
		\includegraphics[width=1.0\textwidth]{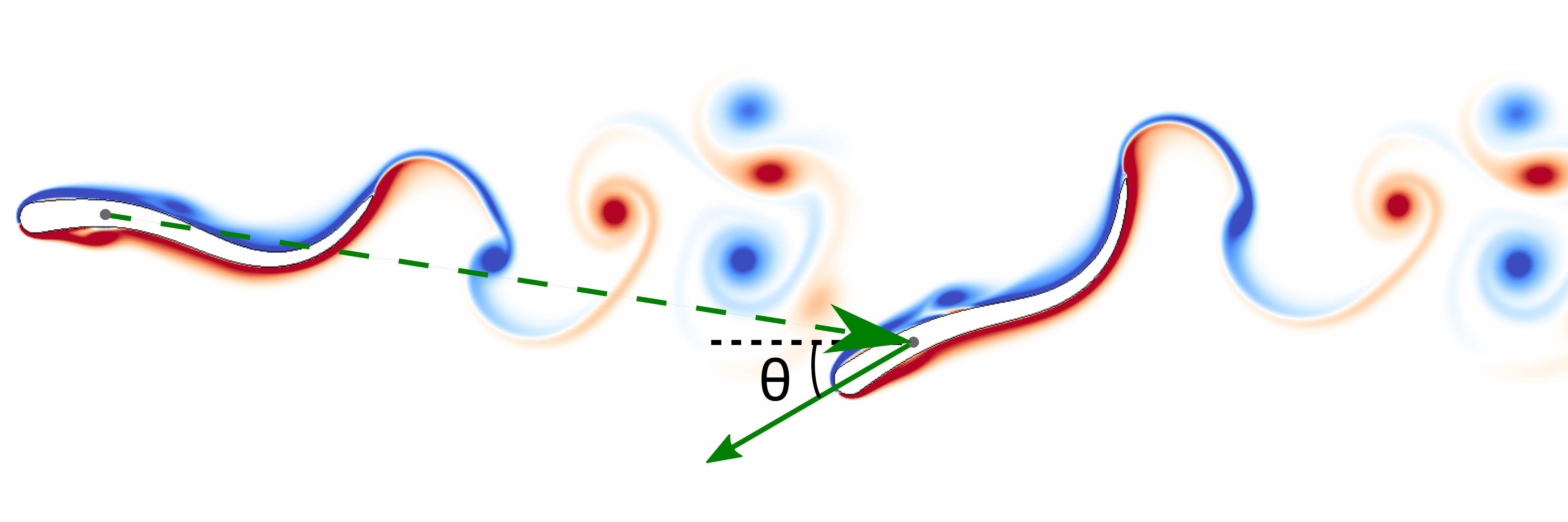}
		\subcaption{}
		\label{fig:states}
	\end{subfigure}
	\caption{\small{\textbf{Fluid-mediated interaction, and visual cues for a smart swimmer.} (\subref{fig:3D2fish}) 3D simulation of two nonautonomous swimmers, in which the leader swims steadily, and the follower maintains a specified relative position such that it interacts favourably with the leader's wake. (\subref{fig:3Dpair}) 3D simulation of three nonautonomous swimmers, where the two followers maintain specified relative positions that are beneficial. The flow-structures have been visualized using iso-surfaces of the Q-criterion. The vortex rings shed by each swimmer spread out in a diverging V-shaped pattern due to their self-induced velocity. An animation of the 3D simulation is provided in \MovieOne. (\subref{fig:goldenRatio}) Comparison of vorticity field in the wake of 2D (top panel) and 3D (bottom panel) swimmers (red: positive, blue: negative). Every half a tail-beat period, the smart-follower in 2D simulations (\Aeta) autonomously selects the most appropriate action encoded in policy \Peta learned during training-simulations, which allows it to maximize long-term swimming-efficiency (\MovieTwo). The smart-follower is capable of adapting to deviations in the leader's trajectory (\MovieThree), as these situations are encountered when performing random actions during training. (\subref{fig:states}) The smart-swimmer relies on a pre-defined set of variables to identify its `observed-state', some of which are depicted in this figure. Additional observed-state parameters are described in the Methods section.}}
\label{fig:configuration}
\end{figure}

During fish propulsion, body undulations and the sideways displacement of  the caudal fin generate and  inject a series of vortex rings in its wake\cite{Muller2001rings,Kern2006,Borazjani2008}. When fish swim in formation, these vortices may assist the locomotion of fish that intercept them judiciously, which in turn can reduce the collective swimming effort. Such vortex-induced benefits have been observed in trout, which curtail muscle usage by capitalizing on energy injected in the flow by obstacles present in streams~\cite{Liao2003Science,Liao2003b}. Here we examine two self-propelled swimmers in a leader/follower arrangement, and investigate the physical mechanisms that lead to energetically beneficial interactions by considering four distinct scenarios. Two of these involve smart-followers that can take autonomous decisions when interacting with a leader's wake, and are referred to as Interacting Swimmers ($\mathit{IS}$) (e.g., the follower in Fig.~\ref{fig:configuration}). Additionally, we consider two distinct Solitary Swimmers ($\mathit{SS}$) that swim in isolation in an unbounded domain. In the case of  interacting swimmers, \Aeta denotes  swimmers that learn the most efficient way of swimming in the leader's wake (without any positional constraints), and acquire a policy \Peta in the process. In turn, swimmer \Ady attempts to minimize lateral deviations from the leader's path, resulting in optimal policy \Pdy. These autonomous swimmers take decisions by virtue of deep RL, using visual cues from their environment (see Fig.~\ref{fig:states}). The Solitary Swimmers \Seta and \Sdy execute actions identical to \Aeta and \Ady, respectively, and serve as `control' configurations to assess how the absence of a leader's wake impacts swimming-energetics.

\section*{\color{black}Reinforcement learning for autonomous swimmers}

\begin{figure}
        \centering
        \begin{subfigure}[b]{0.49\textwidth}
                \centering
                \includegraphics{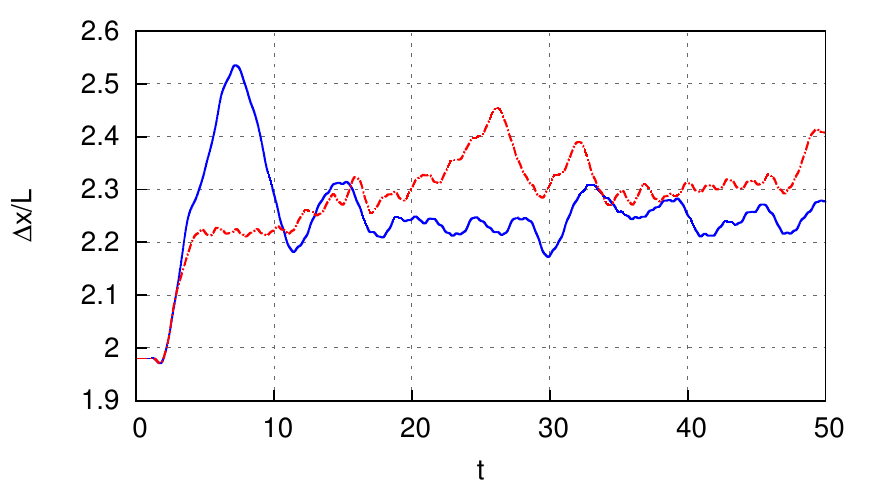}
                \subcaption{}
                \label{fig:dX}
        \end{subfigure}
        \begin{subfigure}[b]{0.49\textwidth}
                \centering
                \includegraphics{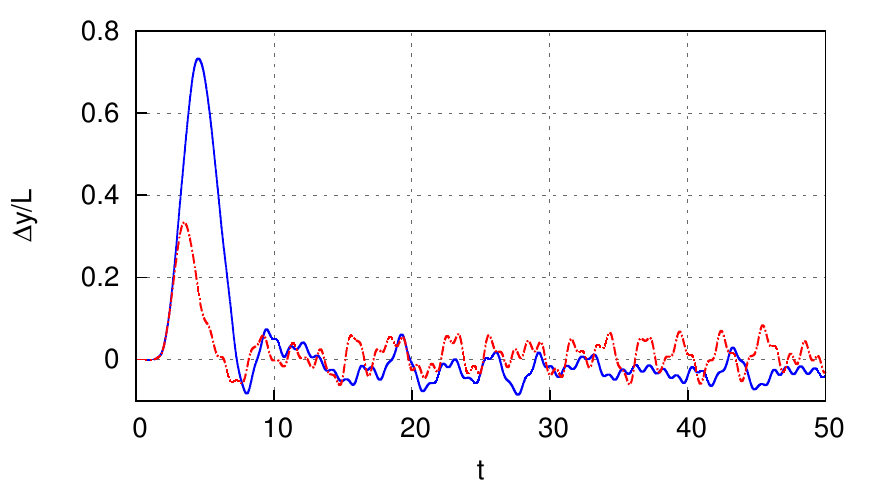}
                \subcaption{}
                \label{fig:dY}
       \end{subfigure}
       \begin{subfigure}[b]{0.49\textwidth}
            	\centering
		\includegraphics[]{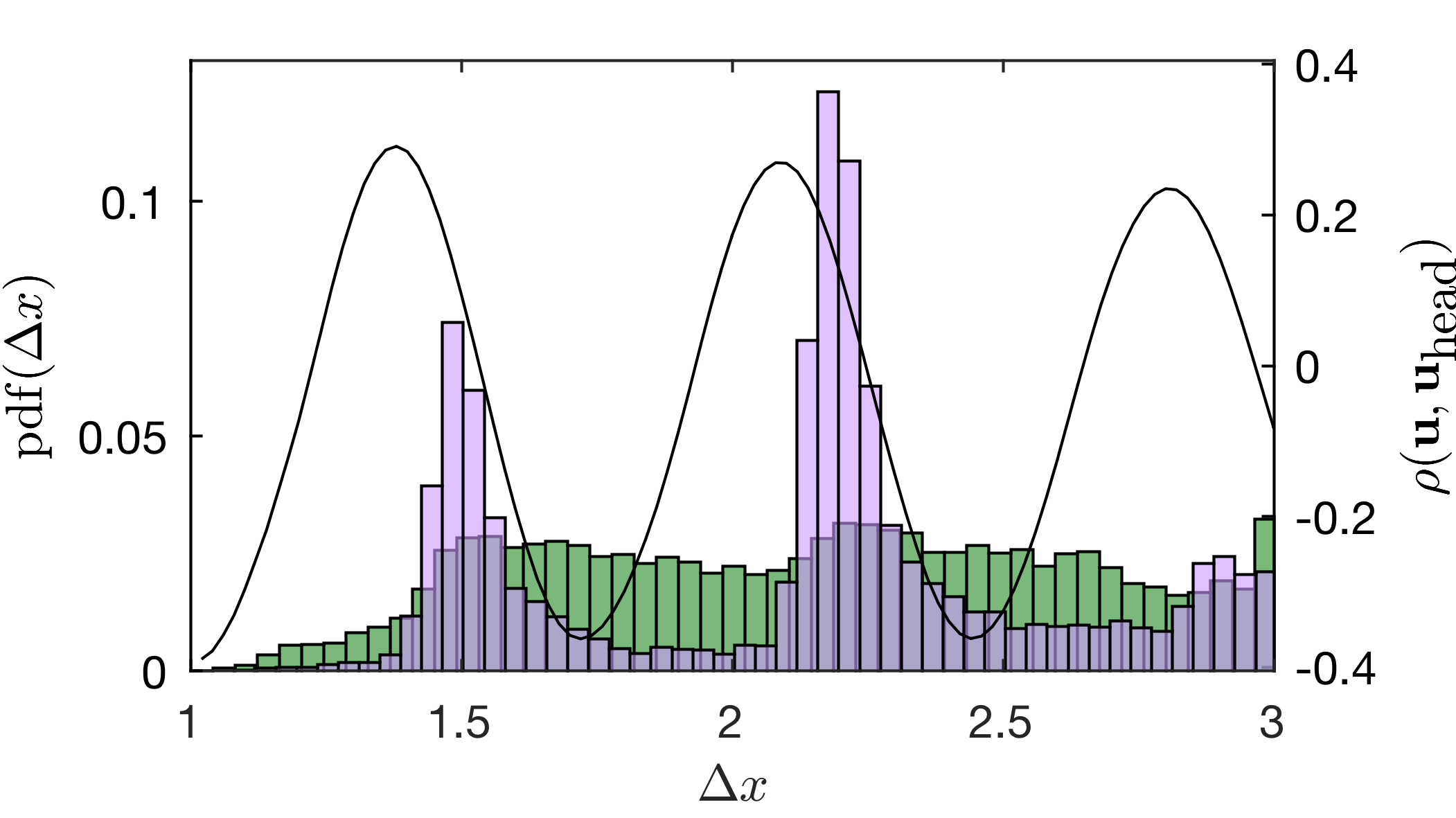}
		\subcaption{}
		\label{fig:histogram}
		\end{subfigure}
        \begin{subfigure}[b]{0.49\textwidth}
            	\centering
		\includegraphics[width=8cm]{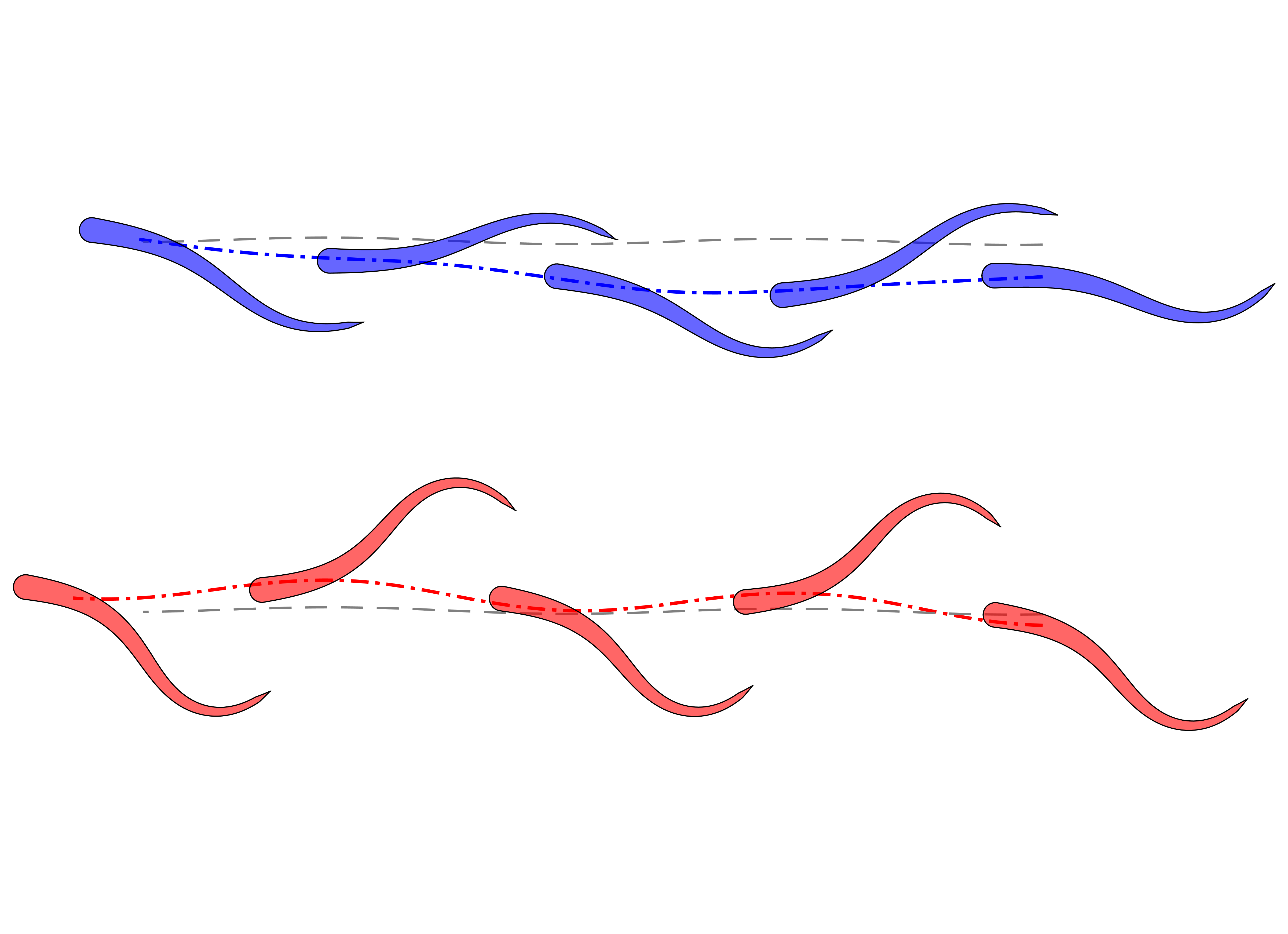}
		\subcaption{}
		\label{fig:contortion}
	\end{subfigure}
		\caption{\textbf{Relative position, correlation with the flow-field, and severity of body-deformation.}  (\subref{fig:dX}) Relative horizontal displacement of the smart followers with respect to the leader, over a duration of 50 tail-beat periods starting from rest (solid blue line - \Aeta, dash-dot red line - \Ady) (\subref{fig:dY}) Lateral displacement of the  smart followers. (\subref{fig:histogram}) Histogram showing the probability density function (pdf - left vertical axis) of swimmer \Aeta's preferred center-of-mass location during training. In the early stages of training (first 10000 transitions - green bars), the swimmer does not show a strong preference for maintaining any particular separation distance. Towards the end of training (last 10000 transitions - lilac bars), the swimmer displays a strong preference for maintaining a separation-distance of either $\Delta x=1.5L$ or $2.2L$. The solid black line in the figure depicts correlation-coefficient, with peaks in the black curve signifying locations where the smart-follower's head-movement would be synchronized with the flow-velocity in an undisturbed wake (please see Supplementary Information for relevant details). (\subref{fig:contortion}) Comparison of body-deformation for swimmers \Aeta (top) and \Ady (bottom), from $t=27$ to $t=29$. Their respective trajectories are shown with the dash-dot lines, whereas the dashed gray line represents the trajectory of the leader (not shown). A quantitative comparison of body-curvature for the two swimmers may be found in Supplementary Fig.~S\ref{fig:contortionQuantitative}.}
\label{fig:posContortion}
\end{figure}

Reinforcement learning ~\cite{Sutton1998}  has been introduced to identify navigation policies in several model systems of  vortex dipoles, soaring birds and micro-swimmers \cite{Gazzola2016, Reddy2016,Colabrese2017}.  Here,  we expand on our earlier work \cite{Gazzola2014,Novati2017} combining Reinforcement Learning with Direct Numerical Simulations of the Navies stokes equations for two self-propelled and autonomous swimmers.  We first investigate  two-dimensional swimmers in a tandem configuration and analyse their kinematics for the cases of \Aeta and \Ady (Fig.~\ref{fig:posContortion}). In both cases, the swimmer trails a leader representing an adult zebrafish of length $L$, swimming steadily at a velocity $U$ (Reynolds number $Re = U L/\nu \approx 5000$). We employ deep Reinforcement Learning (see Methods section for details), and after training we observe that \Ady is able to maintain its position behind the leader quite effectively ($\Delta y\approx0$, Fig.~\ref{fig:dY}), in accordance to its reward  ($R_{d} =1-\lvert\Delta y\rvert/L$). Surprisingly, \Aeta with a reward function proportional to swimming-efficiency ($R_{\eta}=\eta$), also settles close to the center of the leader's wake (Fig.~\ref{fig:dY} and \MovieTwo), although it receives no reward associated with its relative position. Both \Ady and \Aeta maintain a distance of $\Delta x\approx 2.2L$ from their respective leaders (Figure~\ref{fig:dX}). \Aeta shows a greater proclivity to maintain this separation and intercepts the periodically shed wake-vortices just after they have been fully formed and detach from the leader's tail. In addition to $\Delta x=2.2L$, there is an additional point of stability at $\Delta x=1.5$ (Fig.~\ref{fig:histogram}). The difference $0.7L$ matches the distance between vortices in the wake of the leader. In both positions the lateral motion of the follower's head is synchronized with the flow-velocity in the leader's wake, thus inducing minimal disturbance on the oncoming flow-field. We note that a similar synchronization has been observed when trout minimize muscle usage by interacting with vortex-columns in a cylinder's wake~\cite{Liao2003Science}. \Aeta undergoes relatively minor body-deformation while manoeuvring (Figure~\ref{fig:contortion}), whereas \Ady executes aggressive turns involving large body-curvature. Trout interacting with cylinder-wakes exhibit increased body-curvature~\cite{Liao2003b}, which is contrary to the behaviour displayed by \Aeta. The difference may be ascribed to the widely-spaced vortex columns generated by large-diameter cylinders used in the experimental study. Weaving in and out of comparatively smaller vortices generated by like-sized fish encountered in a school (Fig.~\ref{fig:goldenRatio}) would entail excessive energy consumption. We note that maintaining $\Delta y=0$ requires significant effort by \Ady (Supplementary Fig.~S\ref{fig:pDefComparison}) since its reward ($R_{d}$) is insensitive to energy expenditure. A previous study~\cite{Novati2017} suggested that minimizing lateral displacement led to enhanced swimming-efficiency (compared to the leader), albeit with noticeable deviation from $\Delta y=0$. In the current study, recurrent neural networks {\color{black} augmented with `Long Short-Term Memory' cells} (Supplementary Fig.~S\ref{fig:NNstructure}) help to encode time-dependencies in the value function, and enable far more robust smart-swimmers. Thus, {\color{black} stringent attempts by \Ady to correct for oscillations about $\Delta y=0$ (Fig.~\ref{fig:dY}) give rise to increased costs (Supplementary Fig.~S\ref{fig:compareSolo}).}

\section*{\color{black} Intercepting vortices for efficient swimming}

\begin{figure*}
        \centering
        \begin{subfigure}[b]{0.49\textwidth}
                \centering
                \includegraphics{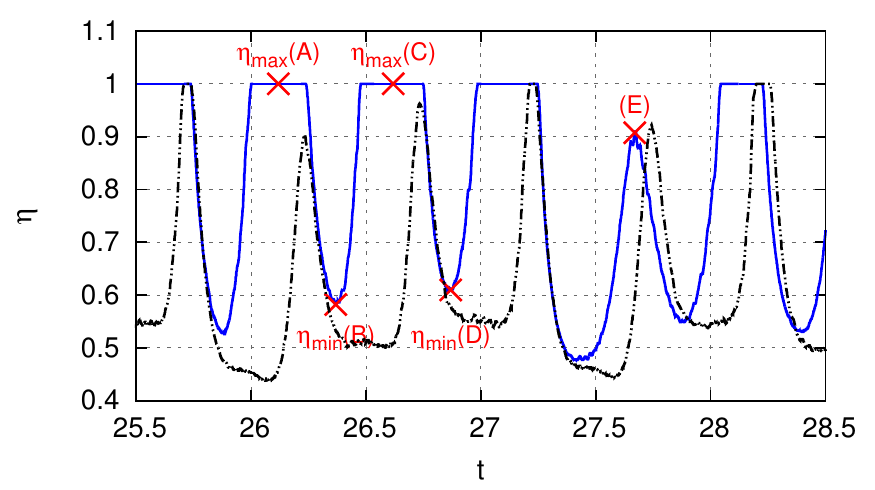}
                \subcaption{}
                \label{fig:etaPlot}
        \end{subfigure}
        \begin{subfigure}[b]{0.49\textwidth}
                \centering
                \includegraphics{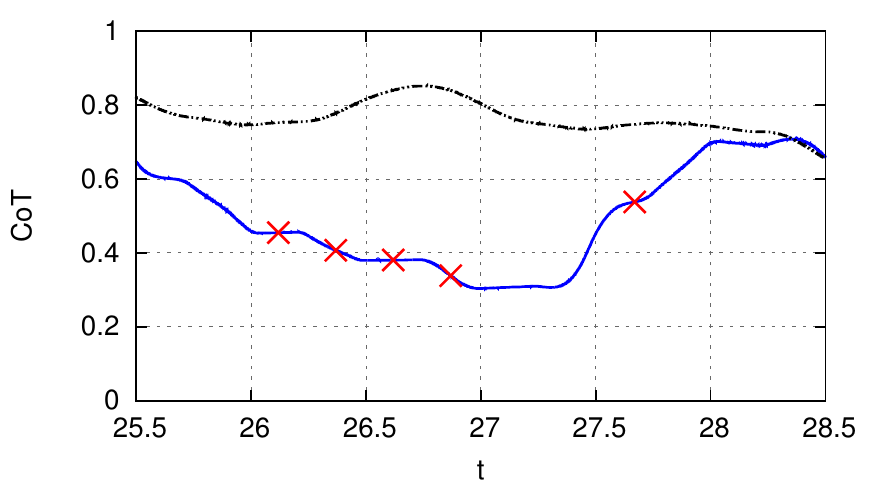}
                \subcaption{}
                \label{fig:cotPlot}
        \end{subfigure}
 		\caption{\textbf{Energetics data for a smart follower maximizing its swimming-efficiency.} (\subref{fig:etaPlot}) Swimming-efficiency, and (\subref{fig:cotPlot}) Cost of Transport for \Aeta (solid blue line) and \Seta (dash-double-dot black line), normalized with respect to the CoT of a steady solitary swimmer. Four instances of maximum and minimum efficiency, which occur periodically throughout the simulation at times $\left(nT_p+0.12\right)$, $\left(nT_p+0.37\right)$, $\left(nT_p+0.62\right)$, $\left(nT_p+0.87\right)$, have been highlighted. $T_p=1$ denotes the constant tail-beat period of the swimmers, whereas $n$ represents an integral multiple. The decline in $\eta$ at point $E$ ($t\approx27.7$, $\eta=0.86$) results from an erroneous manoeuvre at $t\approx26.5$ (\MovieFour), which reveals the existence of a time-delay between actions and their consequences.}
\label{fig:rEtaData}
\end{figure*}

To determine the impact of wake-induced interactions on swimming-performance, we compare energetics data for \Aeta and \Seta (Fig.~\ref{fig:rEtaData}). The swimming-efficiency of \Aeta is significantly higher than that of \Seta (Fig.~\ref{fig:etaPlot}), whereas the Cost of Transport (CoT), which represents energy spent for traversing a unit distance, is lower (Fig.~\ref{fig:cotPlot}). Over a duration of 10 tail-beat periods (from $t=20$ to $t=30$, Supplementary Fig.~S\ref{fig:compareSolo}) \Aeta experiences a $11\%$ increase in average speed compared to \Seta, a $32\%$ increase in average swimming-efficiency, and a $36\%$ decrease in CoT. The benefit for \Aeta results from both a $29\%$ reduction in effort required for deforming its body against flow-induced forces ($P_{Def}$), and a $53\%$ increase in average thrust-power ($P_{Thrust}$). Performance-differences between \Aeta and \Seta exist solely due to the presence/absence of a preceding wake, since both swimmers undergo identical body-undulations throughout the simulations. Comparing the swimming-efficiency and power values of four distinct swimmers (Supplementary Fig.~S\ref{fig:compareSolo} and Supplementary Table~\ref{tab:compare}), we confirm that \Aeta and \Seta are considerably more energetically efficient than either \Ady or \Sdy, thus verifying the hydrodynamic benefits of coordinated swimming.

\begin{figure*}
        \centering
        \begin{subfigure}[b]{0.59\textwidth}
                \centering
                \includegraphics[width=1.0\textwidth]{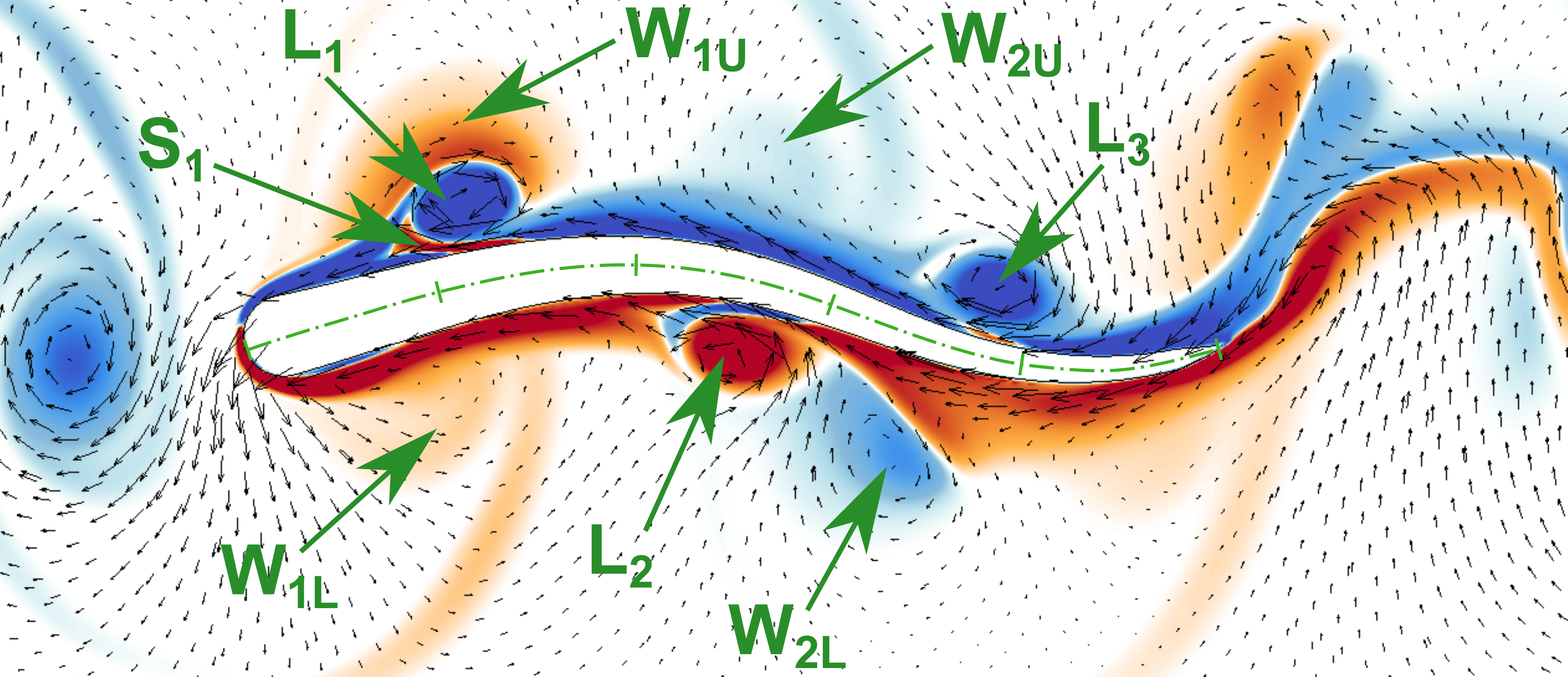}
                 \includegraphics[width=1.0\textwidth]{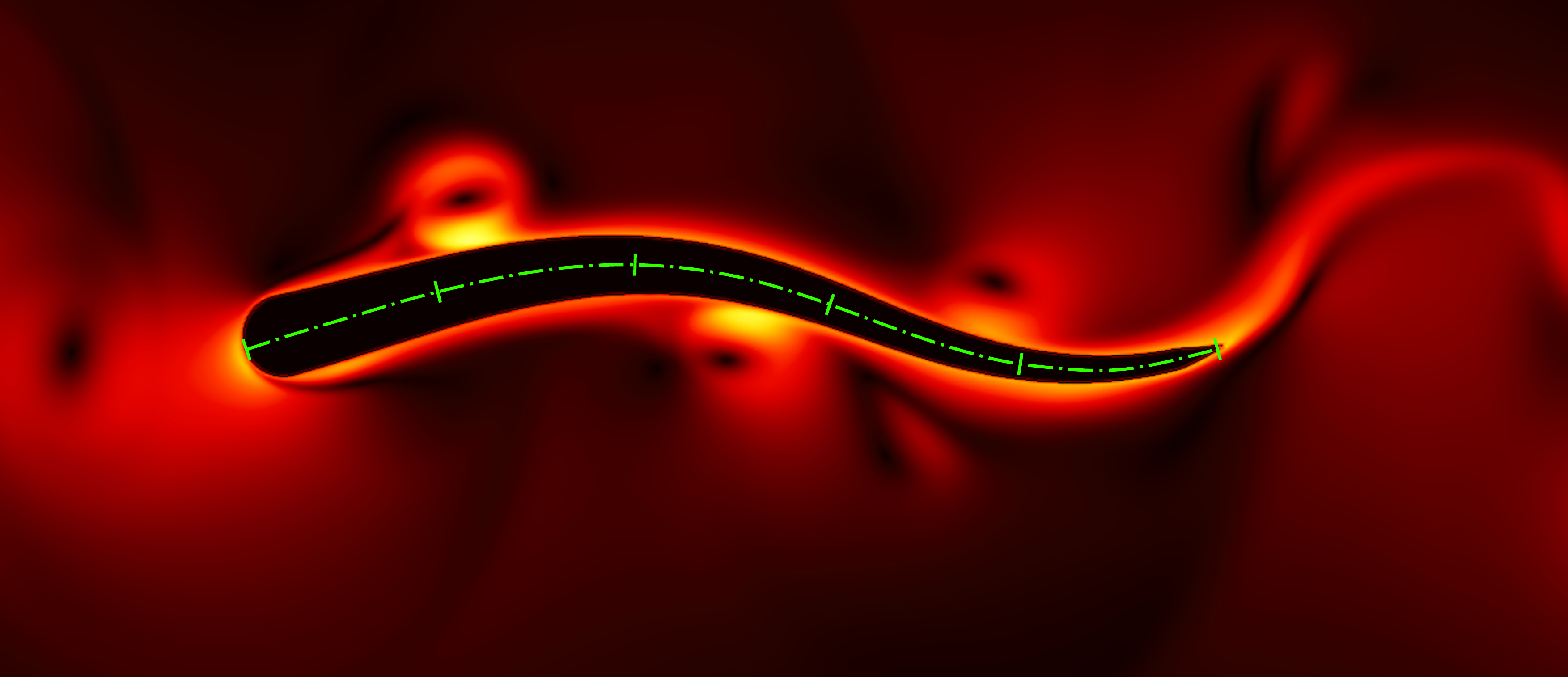}
                \subcaption{}
                \label{fig:vortVelEtaMaxA}
        \end{subfigure}
        \begin{subfigure}[b]{0.39\textwidth}
                \centering
                \includegraphics{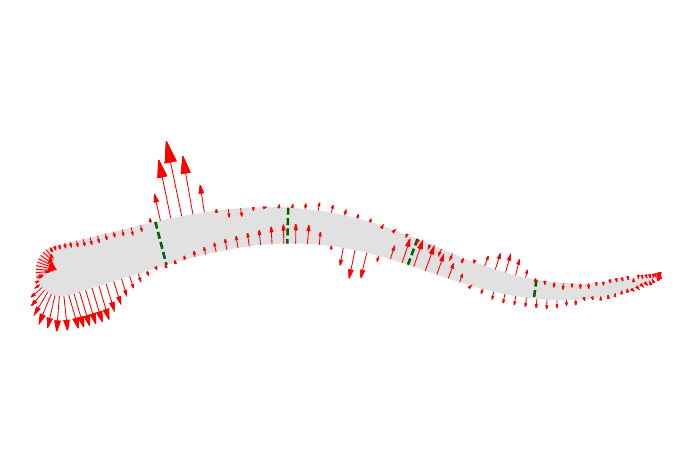}
                \includegraphics{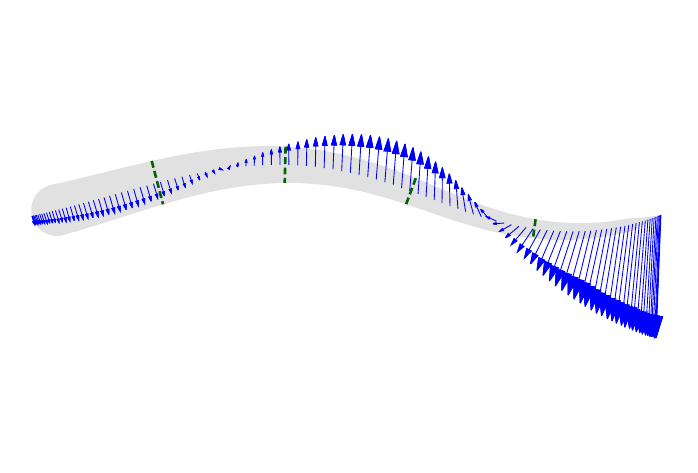}
                \subcaption{}
                \label{fig:kinematicsEtaMaxA}
        \end{subfigure} \\
 		\begin{subfigure}[b]{0.49\textwidth}
                \centering
                \includegraphics{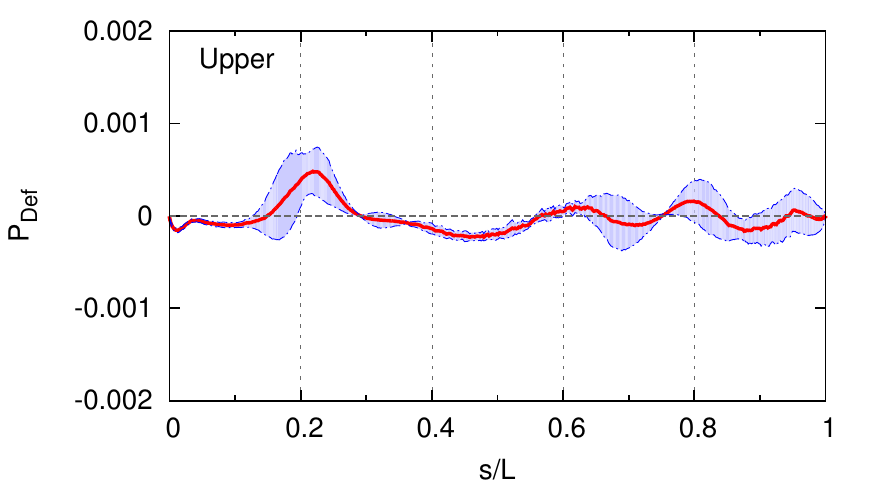}
                \subcaption{}
                \label{fig:pDefUEtaMaxA}
        \end{subfigure}
        \begin{subfigure}[b]{0.49\textwidth}
                \centering
                \includegraphics{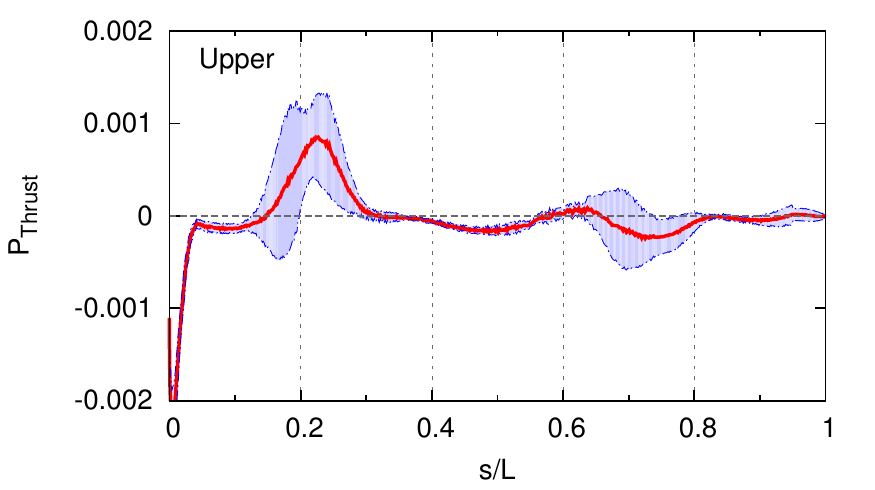}
                \subcaption{}
                \label{fig:pThrustUEtaMaxA}
        \end{subfigure} \\
        \begin{subfigure}[b]{0.49\textwidth}
                \centering
                \includegraphics{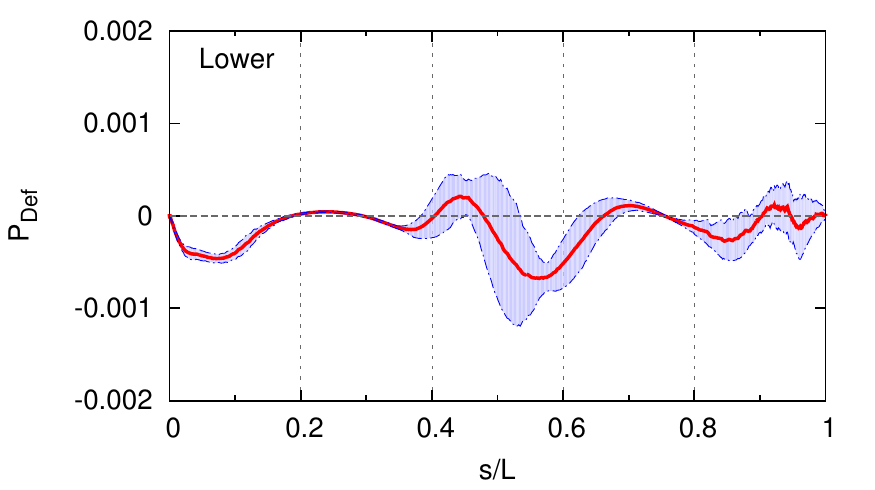}
                \subcaption{}
                \label{fig:pDefLEtaMaxA}
        \end{subfigure}
        \begin{subfigure}[b]{0.49\textwidth}
                \centering
                \includegraphics{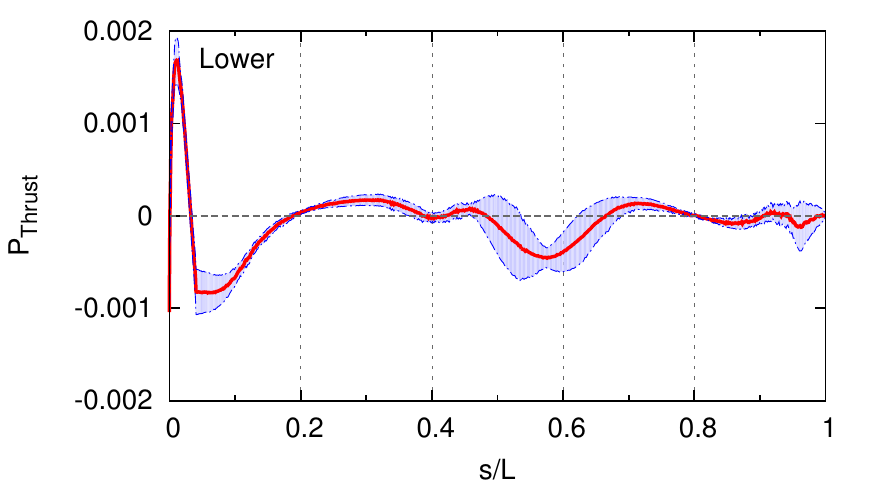}
                \subcaption{}
                \label{fig:pThrustLEtaMaxA}
        \end{subfigure}
  \caption{\textbf{Flow-field, and flow-induced forces for \Aeta, corresponding to maximum efficiency.} \smaller{(\subref{fig:vortVelEtaMaxA}) Vorticity field (red: positive, blue: negative) with velocity vectors shown as black arrows (top), and velocity magnitude shown in the lower panel (bright: high speed, dark: low speed). The snapshots correspond to $t=26.12$, i.e., point $\eta_{max}(A)$ in Fig.~\ref{fig:etaPlot}. In all the panels, demarcations are shown at every $0.2L$ along the body center-line for reference. The wake-vortices intercepted by the follower ($W_{1U}$, $W_{1L}$, $W_{2U}$, $W_{2L}$), the lifted-vortices created by interaction of the body with the flow ($L_1$, $L_2$, and $L_3$), and secondary-vorticity $S_1$ generated by $L_1$ have been annotated in the figure. (\subref{fig:kinematicsEtaMaxA}) Flow-induced force-vectors (top) and body-deformation velocity (bottom) at $t=26.12$. (\subref{fig:pDefUEtaMaxA}) Deformation-power, and (\subref{fig:pThrustUEtaMaxA}) thrust-power (with negative values indicating drag-power) acting on the upper surface of follower. The red line indicates the average over 10 different snapshots ranging from $t=30.12$ to $t=39.12$. The envelope signifies the standard deviation among the 10 snapshots. (\subref{fig:pDefLEtaMaxA}) Deformation-power and (\subref{fig:pThrustLEtaMaxA}) thrust-power on the lower (left-lateral) surface of the swimmer.}}
\label{fig:fluidsEtaMaxA}
\end{figure*}

The efficient swimming of \Aeta (e.g., point $\eta_{max}(A)$ in Fig.~\ref{fig:etaPlot}) is attributed to the  synchronized  motion of its head with the lateral flow-velocity generated by the wake-vortices of the leader (see panel `v' in \MovieTwo). This mechanism is evidenced by the correlation-curve shown in Fig.~\ref{fig:histogram}, and by the co-alignment of velocity vectors close to the head in Figs.~\ref{fig:vortVelEtaMaxA} and~\ref{fig:kinematicsEtaMaxA}. As shown in \MovieFour, \Aeta intercepts the oncoming vortices in a slightly skewed manner, splitting each vortex into a stronger ($W_{1U}$, Fig.~\ref{fig:vortVelEtaMaxA}) and a weaker fragment ($W_{1L}$). The vortices interact with the swimmer's own boundary layer to generate `lifted-vortices' ($L_1$), which in turn generate secondary-vorticity ($S_1$) close to the body. Meanwhile, the wake- and lifted-vortices created during the previous half-period, $W_{2U}$, $W_{2L}$, and $L_{2}$, have travelled downstream along the body. This sequence of events alternates periodically between the upper (right-lateral) and lower (left-lateral) surfaces, as seen in \MovieFour. Interactions of \Aeta with the flow-field at points $\eta_{min}(D)$ and $(E)$ in Fig.~\ref{fig:etaPlot} are analyzed separately in Supplementary Figs.~S\ref{fig:fluidsPointEtaMinD} and S\ref{fig:examinePointE}.


We  observe that the swimmer's upper surface is covered in a layer of negative vorticity (and vice versa for the lower surface) (Fig.~\ref{fig:vortVelEtaMaxA}, top panel) owing to the no-slip boundary condition. The  wake- or the lifted-vortices weaken this distribution  by generating vorticity of opposite sign (e.g., secondary-vorticity visible in narrow regions between the fish-surface and vortices $L_1$, $W_{1L}$, $L_2$, and $L_3$), and create high-speed areas visible as bright spots in Fig.~\ref{fig:vortVelEtaMaxA} (lower panel). The resulting low-pressure region exerts a suction-force on the surface of the swimmer (Fig.~\ref{fig:kinematicsEtaMaxA}, upper panel), which assists body-undulations when the force-vectors coincide with the deformation-velocity (Fig.~\ref{fig:kinematicsEtaMaxA} lower panel), or increases the effort required when they are counter-aligned. The detailed impact of these interactions is demonstrated in  Figs.~\ref{fig:pDefUEtaMaxA} to~\ref{fig:pThrustLEtaMaxA}. On the lower surface, $W_{1L}$ generates a suction-force oriented in the same direction as the deformation-velocity ($0<s<0.2L$ in Fig.~\ref{fig:kinematicsEtaMaxA}), resulting in negative $P_{Def}$ (Fig.~\ref{fig:pDefLEtaMaxA}) and favourable $P_{Thrust}$ (Fig.~\ref{fig:pThrustLEtaMaxA}). On the upper surface, the lifted-vortex $L_1$ increases the effort required for deforming the body (positive peak in Fig.~\ref{fig:pDefUEtaMaxA} at $s=0.2L$), but is beneficial in terms of producing large positive thrust-power (Fig.~\ref{fig:pThrustUEtaMaxA}). Moreover, as $L_1$ progresses along the body, it results in a prominent reduction in $P_{Def}$ over the next half-period, similar to the negative peak produced by the lifted-vortex $L_2$ ($s=0.55L$ in Fig.~\ref{fig:pDefLEtaMaxA}). The average $P_{Def}$ on both the upper and lower surfaces is predominantly negative (i.e., beneficial), in contrast to the minimum swimming-efficiency instance $\eta_{min}(D)$, where a mostly positive $P_{Def}$ distribution signifies substantial effort required for deforming the body (Supplementary Fig.~S\ref{fig:fluidsPointEtaMinD}). We observe noticeable drag on the upper surface close to $s=0$ (Fig.~\ref{fig:kinematicsEtaMaxA} top panel and Fig.~\ref{fig:pThrustUEtaMaxA}), attributed to  high-pressure region  forming in front of the swimmer's head. Forces induced by $W_{1L}$ are both beneficial and detrimental in terms of generating thrust-power ($0<s<0.2L$ in Fig.~\ref{fig:pThrustLEtaMaxA}), whereas forces induced by $L_2$ primarily increase drag but assist in body-deformation (Fig.~\ref{fig:pDefLEtaMaxA}). The tail-section ($s=0.8L$ to $1L$) does not contribute noticeably to either thrust- or deformation-power at the instant of maximum swimming-efficiency.

\section*{\color{black}Energy-saving mechanisms in coordinated swimming}


The most discernible behaviour of \Aeta is the synchronization of its head-movement with the wake-flow.  However, the most prominent reduction in deformation-power occurs near the midsection of the body ($0.4\le s \le 0.7$ in Figs.~\ref{fig:pDefUEtaMaxA} and~\ref{fig:pDefLEtaMaxA}). This indicates that the technique devised by \Aeta is markedly different from energy-conserving mechanisms implied in previous theoretical~\cite{Weihs1973,Weihs1975} and computational~\cite{Daghooghi2015} work, namely, drag-reduction attributed to reduced relative-velocity in the flow, and thrust-increase owing to the `chanelling effect'. In fact, the predominant energetics-gain (i.e., negative $P_{Def}$) occurs in areas of high relative-velocity, for instance near the high-velocity spot generated by vortex $L_2$ (Fig.~\ref{fig:fluidsEtaMaxA}). This dependence of swimming-efficiency on a complex interplay between wake-vortices and body-deformation aligns closely with experimental findings~\cite{Liao2003Science,Liao2003b}.

\begin{figure*}
        \centering
        \begin{subfigure}[b]{0.49\textwidth}
                \centering
                 \includegraphics[width=\textwidth]{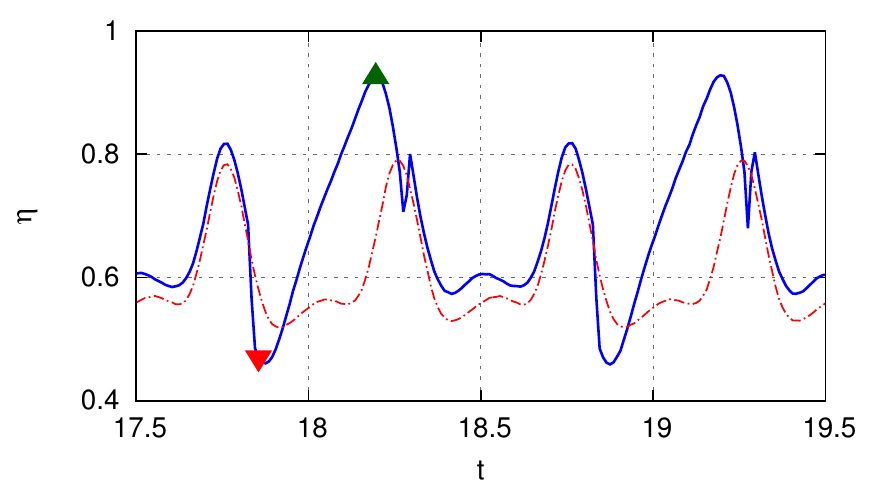}
                \subcaption{}
                \label{fig:eta3DPlot}
        \end{subfigure}
         \begin{subfigure}[b]{0.49\textwidth}
                \centering
                {\setlength{\fboxsep}{0pt}
                \fbox{\includegraphics[width=\textwidth]{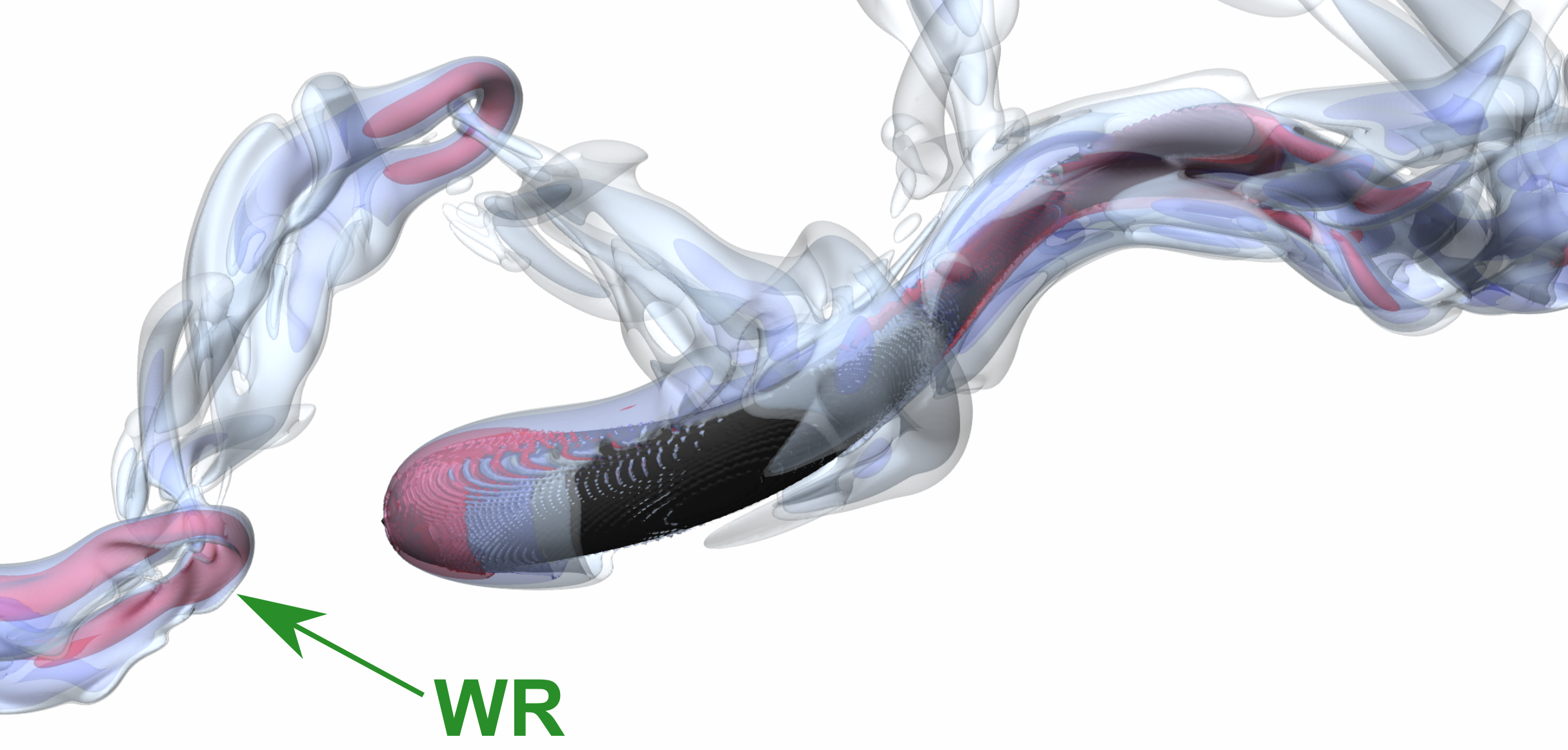}}
                }
                \subcaption{}
                \label{fig:3DframeApproach}
        \end{subfigure}
        \begin{subfigure}[b]{0.49\textwidth}
                \centering
               {\setlength{\fboxsep}{0pt}
                \fbox{\includegraphics[width=\textwidth]{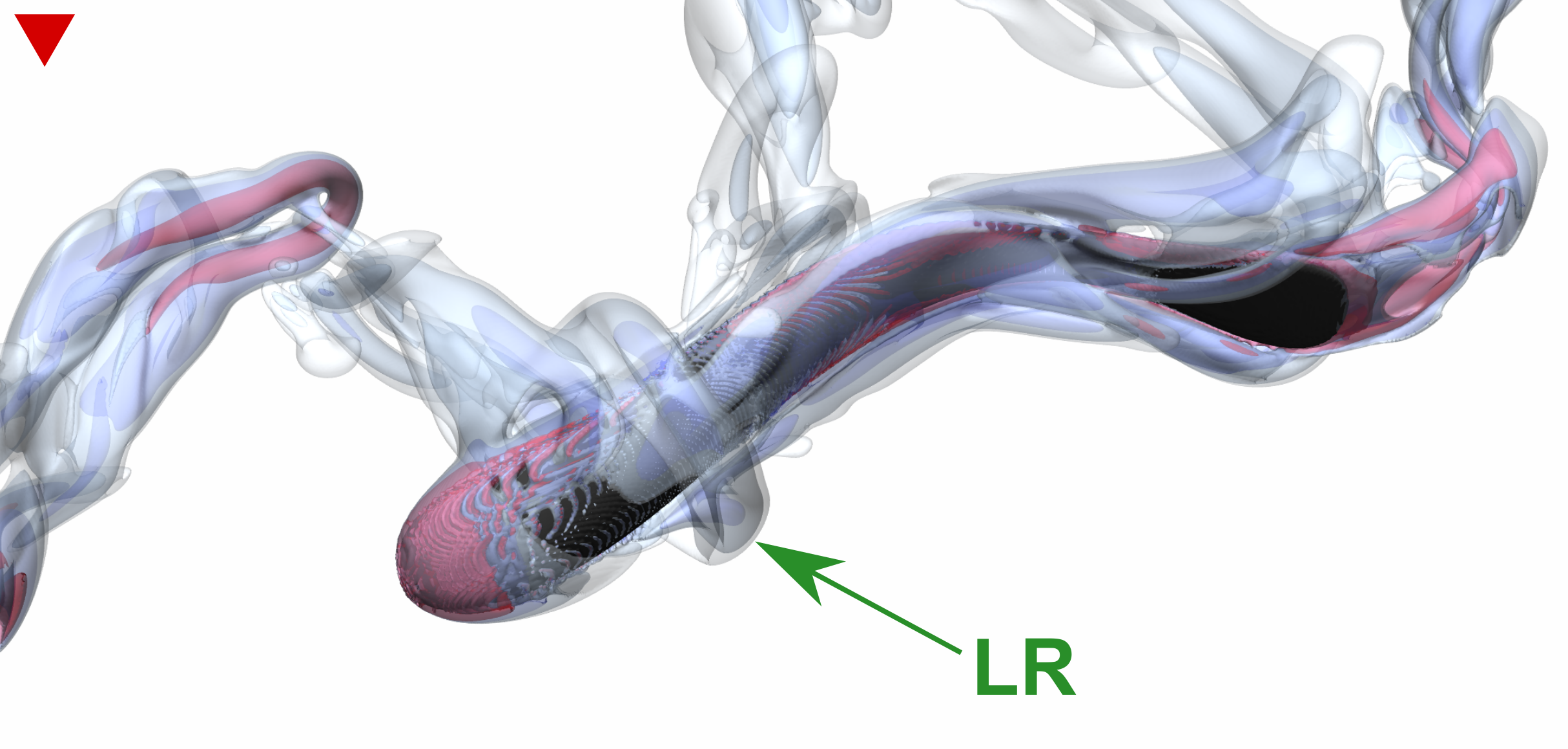}}
                }
                \subcaption{}
                \label{fig:3DframeBad}
        \end{subfigure}
        \begin{subfigure}[b]{0.49\textwidth}
                \centering
                {\setlength{\fboxsep}{0pt}
                \fbox{\includegraphics[width=\textwidth]{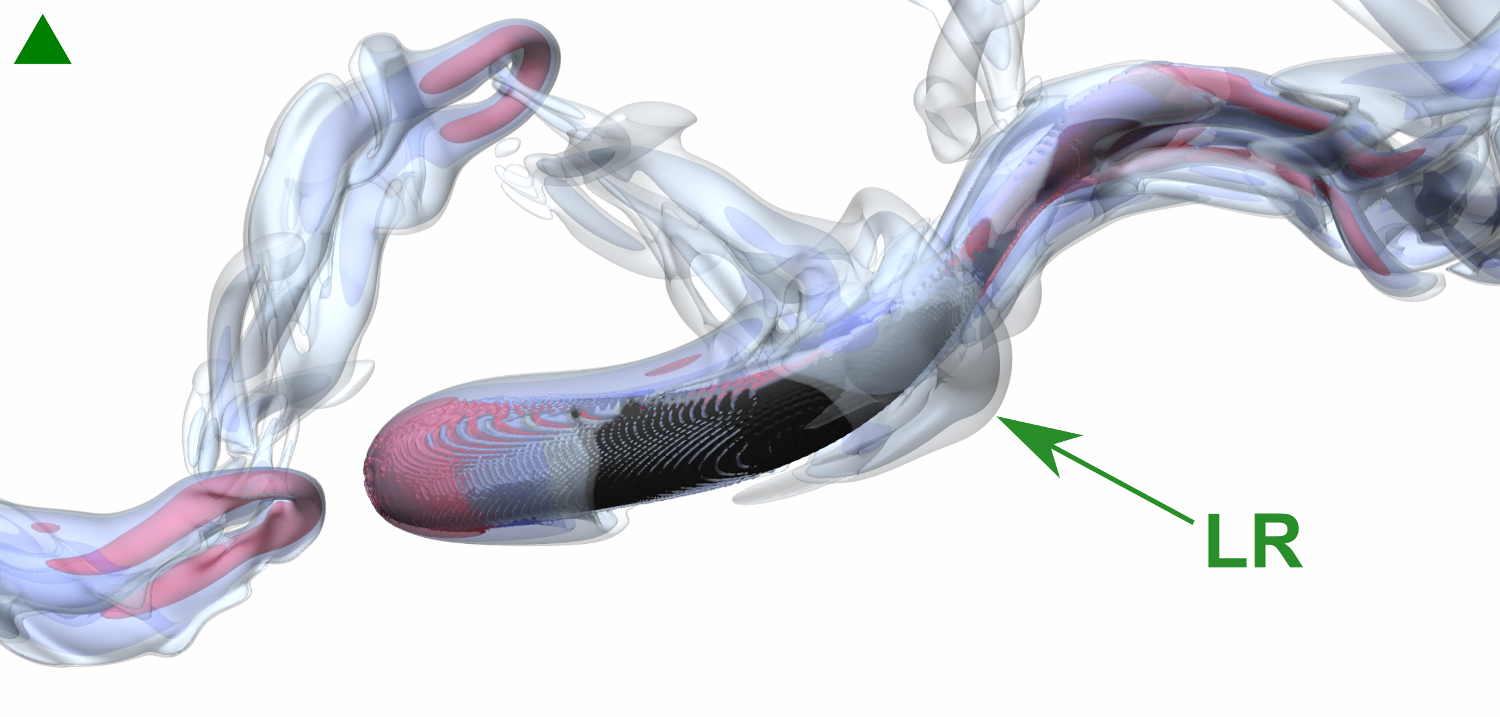}}
                }
                \subcaption{}
                \label{fig:3DframeGood}
        \end{subfigure}
 		\caption{\textbf{3D swimmer interacting with wake vortex rings.} (\subref{fig:eta3DPlot}) Swimming-efficiency for a three-dimensional leader (dash-dot red line) and a follower (solid blue line) that adjusts its undulations via a Proportional-Integrator (PI) feedback controller to maintain a specified position in the wake. After an initial transient, the patterns visible in the efficiency-curves repeat periodically with $T_p$. Time-instances where the follower attains its minimum and maximum swimming-efficiency have been marked with an inverted red triangle, and an upright green triangle, respectively. The sudden jumps at $t\approx 18.3$ and $19.3$ correspond to adjustments made by the PI controller. (\subref{fig:3DframeApproach}) An oncoming wake-vortex ring (WR) is intercepted by the head of the follower, and generates a new `lifted-vortex' ring (LR, panel \subref{fig:3DframeBad}) similar to the 2D case (Fig.~\ref{fig:fluidsEtaMaxA}). As this ring interacts with the deforming body, it lowers the swimming-efficiency initially ($t\approx 17.8$: panels \subref{fig:eta3DPlot} and \subref{fig:3DframeBad}), but provides a noticeable benefit further dowstream ($t\approx 18.2$, panels \subref{fig:eta3DPlot} and \subref{fig:3DframeGood}).}
\label{fig:3Dsnapshots}
\end{figure*}

We remark that the majority of the results presented here were obtained with a steadily-swimming leader. However, with no additional training, \Aeta is able to extract an energetic-benefit even when exposed to an erratic leader (as seen in \MovieThree), where it deliberately chooses to interact with the unsteady wake. Moreover, given the head-synchronization tendency of the 2D smart-swimmer, we identify suitable locations behind a 3D leader where the flow velocity would match a follower's head motion (Supplementary Fig.~S\ref{fig:correlation3D}). A feedback controller is used to regulate the undulations of two followers to maintain these target coordinates on either branch of the diverging wake, as shown in Fig.~\ref{fig:3Dpair} and \MovieOne. The controlled motion yields an $11\%$ increase in average swimming-efficiency for each of the followers (Fig.~\ref{fig:eta3DPlot}), and a $5\%$ reduction in each of their Cost of Transport. Overall, the group experiences a $7.4\%$ increase in efficiency when compared to three isolated non-interacting swimers. The mechanism of energy-savings closely resembles that observed for the 2D swimmer; an oncoming wake-vortex ring (WR - Fig.~\ref{fig:3DframeApproach}) interacts with the deforming body to generate a `lifted-vortex' ring (LR - Fig.~\ref{fig:3DframeBad}). As this new ring proceeds along the length of the body, it modulates the follower's swimming-efficiency as observed in Fig.~\ref{fig:3Dsnapshots}. Remarkably, the positioning of the lifted-ring at the instants of minimum and maximum swimming-efficiency resembles the corresponding positioning of lifted-vortices in the 2D case; a slight dip in efficiency corresponds to lifted-vortices interacting with the anterior section of the body (Fig.~\ref{fig:3DframeBad} and Supplementary Fig.~S\ref{fig:fluidsPointEtaMinD}), whereas an increase occurs upon their interaction with the midsection (Fig.~\ref{fig:3DframeGood} and Fig.~\ref{fig:fluidsEtaMaxA}). 

These results showcase the remarkable capability of machine learning, and deep RL  in particular,  for discovering effective solutions that may not have been envisaged by humans, either owing to pre-existing biases, or due to the difficulty of anticipating the effects of delayed reactions by swimmers in complex flows.
Finally,  this study demonstrates that deep reinforcement learning can produce navigation algorithms for complex flow-fields, with promising implications for energy savings in autonomous robotic swarms.




{\textbf{Methods} We perform two-and three dimensional simulations of multiple self-propelled swimmers using wavelet adapted vortex methods \cite{Rossinelli2015} to discretise the velocity-vorticity form of the Navier-Stokes (NS) equations (in 2D), and their velocity-pressure form along with the pressure-projection \cite{Chorin1968} method (in 3D) using finite differences on a uniform computational grid. The body-geometry of the self-propelled swimmers is  based on simplified models of a zebrafish. The swimmers adapt their motion using deep reinforcement learning. The learning process was greatly accelerated by employing  recurrent neural networks with long-short term memory (RL-LSTM) \cite{Hochreiter1997} as a surrogate of the value function for the smart-swimmer. Additional details regarding the simulation methods and the reinforcement learning algorithm are provided in the Supporting Information.}

\textbf{Acknowledgements} This work was supported by the European Research Council Advanced Investigator Award (Fluid Mechanics of Collective Behavior, Grant: 341117), and the Swiss National Science Foundation Sinergia Award (CRSII3 147675). The authors are grateful to the Swiss National Supercomputing Center (CSCS) for providing access to computational resources (project `s658').

\bibliography{deepRLfish}

\bibliographystyle{pnas-new}

 





\newpage

\section*{Supporting Information - Methods}
\paragraph*{Simulation details.}
The simulations presented here are based on the incompressible Navier-Stokes (NS) equations:
\begin{eqnarray}
\nabla\cdot\bm{u} &=& 0\\
\dfrac{\partial \bm{u}}{\partial t}+ \bm{u}\cdot\nabla\bm{u} &=& -\dfrac{\nabla P}{\rho_f} + \nu \nabla^2 \bm{u} + \lambda \chi (\bm{u}_s - \bm{u})
\end{eqnarray}
Each swimmer is represented on the computational grid via the characteristic function $\chi$, and interacts with the fluid by means of the penalty~\cite{Coquerelle2008} term $\lambda\chi\left(\bm{u}_s - \bm{u}\right)$, with $\lambda=1e6$. $\bm{u}_s$ denotes the swimmer's combined translational, rotational, and deformation velocity, whereas $\bm{u}$ and $\nu$ correspond to the fluid velocity and viscosity, respectively. $P$ represents the pressure, and the fluid density is denoted by $\rho_f$.

The vorticity form of the NS equations was used for the two-dimensional simulations. A wavelet adaptive grid~\cite{Rossinelli2015} with an effective resolution of $4096^2$ points was used to discretize a unit square domain. A lower effective resolution of $1024^2$ points was used for the training-simulations to minimize computational cost. The pressure-Poisson equation ($\nabla^2P=-\rho_f \left(\nabla\bm{u}^T:\nabla\bm{u}\right) + \rho_f\lambda\nabla\cdot  \left(\chi\left(\bm{u}_s - \bm{u}\right)\right)$), necessary for estimating the distribution of flow-induced forces on the swimmers' bodies, was solved using the Fast Multipole Method~\cite{Greengard1987,Verma2017}.

The three-dimensional simulations employed the pressure-projection method for solving the NS equations~\cite{Chorin1968}. The simulations were parallelized via the CUBISM framework~\cite{Rossinelli2015}, and used a uniform grid consisting of $2048\times1024\times256$ points in a domain of size $1\times0.5\times0.125$. The non-divergence-free deformation of the self-propelled swimmers was incorporated into the pressure-Poisson equation as follows:
\begin{equation}
\nabla^2 P = \dfrac{\rho_f}{\Delta t} \left( \nabla \cdot \bm{u}^\star - \chi \nabla \cdot \bm{u}_s \right),
\label{eq:pressProjection}
\end{equation}
where $\bm{u}^\star$ represents the intermediate velocity from the convection-diffusion-penalization fractional steps. Equation~\ref{eq:pressProjection} was solved using a distributed Fast Fourier Transform library (AccFFT~\cite{Gholami2015}).

\paragraph*{Flow-induced forces, and energetics variables.}
The pressure-induced and viscous forces acting on the swimmers are computed as follows~\cite{Verma2017}:
\begin{eqnarray}
\bm{dF}_P &=& -P\bm{n} \ dS \\
\label{eq:pressForce}
\bm{dF}_\nu &=& 2\mu\bm{D}\cdot \bm{n} \ dS
\label{eq:viscousForce}
\end{eqnarray}
Here, $P$ represents the pressure acting on the swimmer's surface, $\bm{D} = \left(\nabla\bm{u} + \nabla\bm{u}^T\right)/2$ is the strain-rate tensor on the surface, and $dS$ denotes the infinitesimal surface area.
Since self-propelled swimmers generate zero net average thrust (and drag) during steady swimming, we determine the instantaneous thrust as follows:
\begin{equation}
\text{Thrust} = \dfrac{1}{2\|\bm{u}\|}\iint \left(\bm{u}\cdot\bm{dF} + \lvert \bm{u}\cdot\bm{dF}\rvert\right),
\end{equation}
where $\bm{dF} = \bm{dF}_P + \bm{dF}_\nu$. Similarly, the instantaneous drag may be determined as:
\begin{equation}
\text{Drag} = \dfrac{1}{2\|\bm{u}\|}\iint \left(\bm{u}\cdot\bm{dF}  - \lvert \bm{u}\cdot\bm{dF}\rvert\right)
\end{equation}
Using these quantities, the thrust-, drag-, and deformation-power are computed as:
\begin{eqnarray}
P_{Thrust} &=& \text{Thrust} \cdot \|\bm{u}\| \\
P_{Drag} &=& -\text{Drag} \cdot \|\bm{u}\| \\
P_{Def} &=& -\iint \bm{u}_{Def}\cdot\bm{dF}
\end{eqnarray}
where $\bm{u}_{Def}$ represents the deformation-velocity of the swimmer's body. The double-integrals in these equations represent surface-integration over the swimmer's body, and yield measurements for time-series analysis. On the other hand, only the integrand is evaluated when surface-distributions of thrust-, drag-, or deformation-power are required (as in Figs.~\ref{fig:pDefUEtaMaxA} to~\ref{fig:pThrustLEtaMaxA}).

The instantaneous swimming-efficiency is based on a modified form of the Froude efficiency proposed in ref.~\cite{Tytell2004}:
\begin{equation}
\eta = \dfrac{P_{Thrust}}{P_{Thrust} + \max(P_{Def},0)}
\end{equation}
To compute both $\eta$ and the Cost of Transport (CoT), we neglect negative values of $P_{Def}$, which can result from beneficial interactions of the smart-swimmer with the leader's wake:
\begin{equation}
CoT(t) = \dfrac{\int_{t-T_p}^t \max(P_{Def},0) dt}{\int_{t-T_p}^t \|\bm{u}\| dt}
\end{equation}
This restriction accounts for the fact that the elastically rigid swimmer may not store energy furnished by the flow, and yields a conservative estimate of potential savings in the CoT. We note that percentage-changes in $P_{Def}$, reported in the main text and the supplementary section, have been computed using this bounded value to avoid overstating any potential benefits.

\paragraph*{Swimmer shape and kinematics.}
The Reynolds number of the self-propelled swimmers is computed as $Re=L^2/\left(\nu T_p\right)$. The body-geometry is based on a simplified model of a zebrafish~\cite{vanRees2013}. The half-width of the 2D profile is described as follows:
\begin{equation}
    w(s)= 
\begin{dcases}
    \sqrt{2w_hs-s^2} 					& 0\le s<s_b\\
    w_h-(w_h-w_t)\left(\dfrac{s-s_b}{s_t-s_b}\right) 	& s_b \le s < s_t \\
    w_t\dfrac{L-s}{L-s_t}				& s_t \le s \le L
\end{dcases}
\label{eq:fishWidth}
\end{equation}
where $s$ is the arc-length along the midline of the geometry, $L=0.1$ is the body length, $w_h=s_b=0.04L$, $s_t=0.95L$, and $w_t=0.01L$. For 3D simulations, the geometry is comprised of elliptical cross sections, with the half-width $w(s)$ and half-height $h(s)$ described via cubic B-splines~\cite{vanRees2013}. Six control-points define the half-width: $(s/L, w/L) = [(0.0, 0.0), \ (0.0, 0.089), \ (1/3, 0.017), \ (2/3, 0.016), \ (1.0, 0.013), \ (1.0, 0.0)]$; whereas eight control-points define the half-height: $(s/L, h/L) = [(0.0, 0.0),\allowbreak \ (0.0, 0.055),\allowbreak \ (0.2, 0.068),\allowbreak \ (0.4, 0.076),\allowbreak \ (0.6, 0.064),\allowbreak \ (0.8, 0.0072),\allowbreak \ (1.0, 0.11),\allowbreak \ (1.0, 0.0)]$. The length was set to $L=0.2$, which keeps the grid-resolution, i.e., the number of points along the fish midline, comparable to the 2D simulations. Body-undulations for both 2D and 3D simulations were generated as a travelling-wave defining the curvature along the midline:
\begin{equation}
k(s,t) = A(s) \sin\left(\dfrac{2\pi t}{T_p} - \dfrac{2\pi s}{L}\right)
\label{eq:curvWave}
\end{equation}
Here $A(s)$ is the curvature amplitude and varies linearly from $A(0)=0.82$ to $A(L)=5.7$.

\paragraph*{Reinforcement Learning.}
Reinforcement learning (RL)~\cite{Sutton1998} is a process by which an agent (in this case, the smart-swimmer) learns to earn rewards through trial-and-error interaction with its environment. At each turn, the agent observes the state of the environment $s_n$ and performs an action $a_n$, which influences both the transition to the next state $s_{n+1}$ and the reward received $r_{n+1}$. The agent's goal is to learn the optimal control policy $a_n=\pi^*(s_n)$ which maximises the action value $Q^*(s_n,a_n)$, defined as the sum of discounted future rewards:
\begin{equation}
Q^* (s_n,a_n) = \max_{\pi}\mathbb{E} \left( r_{n+1} + \gamma r_{n+2}  + \gamma^2 r_{n+3} + \dots  \vert \, a_{m} = \pi(s_m) ~ \forall m \in [n+1, \mathcal{T}] \right)
\end{equation}
Here, $\mathcal{T}$ denotes the terminal state of a training-simulation, and the discount factor $\gamma$ is set to 0.9. The optimal action-value function $Q^*(s_n,a_n)$ is a fixed point of the Bellman equation: $Q^*(s_n,a_n) = \mathbb{E} \left[ r_{n+1} +\gamma \max_{a'} Q^*(s_{n+1},a')\right]$~\cite{Bellman2010}. We approximate $Q^*(s_n,a_n)$ using a neural network~\cite{van2015deep,mnih2015human,riedmiller2005neural} with weights $w_k$, which are updated iteratively to minimize the temporal difference error:
\begin{equation}
\text{TD$_{\text{err}}$} = \mathbb{E}_{s_n, a_n, s_{n+1}} \left[ r_{n+1} + \gamma Q(s_{n+1}, a'; \textrm{w}_-) - Q(s_n,a_n;\textrm{w}_k)\right]
\end{equation}
Here, $w_-$ is a set of target weights, and $a'$ is the best action in state $s_{n+1}$ computed with the current weights ($a' = \argmax_{a} Q(s_{n+1}, a; \textrm{w}_k)$). The target weights $w_-$ are updated towards the current weights as $\textrm{w}_- \gets (1-\alpha) \textrm{w}_- + \alpha \textrm{w}_{k}$, where $\alpha = 10^{-4}$ is an under-relaxation factor used to stabilize the algorithm~\cite{mnih2015human}.

\paragraph*{States and actions.}
The six observed-state variables perceived by the learning agent include $\Delta x$, $\Delta y$, $\theta$, the two most recent actions taken by the agent, and the current tail-beat `stage' $\text{mod}(t,T_p)/T_p$. The permissible range of the observed-state variables is limited to: $1\le\Delta x/L\le 3$; $\lvert\Delta y\lvert/L \le 1$ (boundary depicted by $R_{end}$ in Supplementary Fig.~S\ref{fig:reward}); and $\lvert\theta\rvert\le\pi/2$. If the agent exceeds any of these thresholds, the training-simulation terminates and the agent receives a terminal reward $R_{end}=-1$.

The smart-swimmer (or agent) is capable of manoeuvering by actively manipulating the curvature-wave travelling down the body. This is accomplished by linearly superimposing a piecewise function on the baseline curvature $k(s,t)$ (equation~\ref{eq:curvWave}):
\begin{equation}
k_{\text{Agent}}(s,t) = k(s,t) + A(s)M\!\left(t,T_p,s,L\right)
\end{equation}
The curve $M\!\left(t,T_p,s,L\right)$ is composed of 3 distinct segments:
\begin{equation}
M\!\left(t,T_p,s,L\right) = \sum_{j=0}^2 b_{n-j} \cdot m \left(\dfrac{t-t_{n-j}}{T_p} - \dfrac{s}{L}\right)
\end{equation}
The curve $m$ is a clamped cubic spline with $m(0)=m'(0)=0$, $m(1/2)=m'(1/2)=0$, and $m(1/4)=1$, $m'(1/4)=0$. $t_n$ represents the time-instance when action $a_n$ is taken, whereas $b_n$ represents the corresponding control-amplitude, which may take five discrete values: $0$, $\pm0.25$, and $\pm0.5$.

\paragraph*{Neural network architecture.} One of the assumptions in RL is that the transition probability to a new state $s_{n+1}$ is independent of the previous transitions, given $s_n$ and $a_n$, i.e.,:
\begin{equation}
p(s_{n+1}\,|\,s_n,a_n) = p(s_{n+1}\,|\,s_n,a_n,\dots,s_0,a_0)
\end{equation}
This assumption is invalidated whenever the agent has a limited perception of the environment. In most realistic cases the agent receives an observation $o_n$ rather than the complete state of the environment $s_n$. Therefore, past observations carry information relevant for future transitions (i.e., $p(o_{n+1}\,|\,o_n,a_n) \neq p(o_{n+1}\,|\,o_n,a_n,\dots,o_0,a_0)$), and should be taken into account in order to make optimal decisions. This operation can be approximated by a Recurrent Neural Network (RNN), which can learn to compute and remember important features in past observations. In this work we approximate the action-value function with a LSTM-RNN~\cite{gers2000learning} composed of three layers of 24 fully connected LSTM cells each, and terminating in a linear layer (Supplementary Fig.~S\ref{fig:NNstructure}). The last layer computes a vector of action-values $\mathbf{q}_{n} = Q(o_{n};y_{n-1},\textrm{w}_k)$ with one component $q_{n}^{(a)}$ for each possible action $a$ available to the agent ($y_{n-1}$ represents the activation of the network at the previous turn).

\paragraph*{Training procedure.}
During training, both the leader and the follower (learning agent) start from rest. The leader swims steadily along a straight line, whereas the follower manoeuvers according to the actions supplied to it. Multiple independent simulations run simultaneously, with each of these sending the current observed-state $o_n$ of the agent to a central processor, and in turn receiving the next action $a_n$ to be performed. The central processor computes $a_n$ using an $\epsilon$-greedy policy (with $\epsilon$ gradually annealed from $1$ to $0.1$) from the most recently updated $Q$ function. Once a training-simulation reaches a terminal state (e.g., the follower hits the boundary labelled $R_{end}$ in Supplementary Fig.~S\ref{fig:reward}), all the messages exchanged between the simulation and the central processor are appended to a training set of sequences $\mathcal{R}$~\cite{lin1993reinforcement}. In the meantime, the network is continually updated by sampling $B$ sequences from the set $\mathcal{R}$, according to algorithm~\ref{algo:rDQN}.
\begin{algorithm} 
\caption{\textbf{Asynchronous recurrent DQN algorithm.}}
\label{algo:rDQN}
{\small
initialize network $\textrm{w}_0$ and target network $\textrm{w}_- =\textrm{w}_0$\;
initialize set of transition sequences $\mathcal{R} = \emptyset$\;
\Repeat{ $Q(o, a; \textrm{w}_k) = Q^*(o, a)$ }{
	$N\gets0$\;
	sample batch of $B$ sequences from $\mathcal{R}$\;
	\myFor{sequence $ j \in [1, \dots, B]$ }{
		$[\mathbf{q}_{j,0},  y_{j,0}] = Q(o_{j,0};\emptyset,\textrm{w}_k)$\;
		\myFor{\text{turns} $ n \in [0, \dots, \mathcal{T}_j]$ }{
			$[\mathbf{q}_{j,n+1}, y_{j,n+1}] = Q(o_{j,n+1}; y_{j,n},\textrm{w}_k)$\;
			$[\tilde{\mathbf{q}}_{j,n+1}, \tilde{y}_{j,n+1}] = Q(o_{j,n+1};y_{j,n},\textrm{w}_-)$\;
			$a'= \argmax_{a}  \left[ q_{j,n+1}^{(a)} \right]$\;
			\myIf{$s_{j,n+1}~ \text{is terminal}$} {
				$e_{j,n}  = r_{j,n+1} -  q_{j,n}^{(a_n)}$\;
			}
			\myElse {}{
				$e_{j,n}  = r_{j,n+1} + \gamma \tilde{q}_{j,n+1}^{(a')} -   q_{j,n}^{(a_n)}$\;
			}
			$N \gets N + 1$\;
		}
	}
	perform BPTT: $\Delta \mathrm{w} = \frac{1}{N} \sum_j \sum_n e_{j,n} \nabla_\mathrm{w} q_{j,n}^{(a_n)}$\;
	update weights $\textrm{w}_{k+1}$ with Adam algorithm \cite{kingma2014}\;
	update target network: $\textrm{w}_- \gets (1-\alpha) \textrm{w}_- + \alpha \textrm{w}_{k+1}$\;
	$k \gets k + 1$\;
}}
\end{algorithm}
The batch gradient $\Delta \textrm{w}$ is computed with back propagation through time (BPTT)~\cite{graves2005framewise}. The network weights are then updated with the Adam stochastic optimization algorithm~\cite{kingma2014}.

\paragraph*{Proportional-Integral feedback controller.}
The PI controller modulates the 3D follower's body-kinematics, which allows it to maintain a specific position ($x_{tgt}$, $y_{tgt}$, $z_{tgt}$) relative to the leader:
\begin{equation}
k(s,t) = \alpha(t) A(s) \left[ \sin\left(\dfrac{2\pi t}{T_p} - \dfrac{2\pi s}{L}\right) + \beta(t)\right]
\end{equation}
The factor $\alpha(t)$ modifies the undulation envelope, and controls the acceleration or deceleration of the follower based on its streamwise distance from the target position:
\begin{equation}
\alpha(t) = 1 + f_1 \left(\frac{x-x_{tgt}}{L}\right)
\end{equation}
The term $\beta(t)$ adds a baseline curvature to the follower's midline to correct for lateral deviations:
\begin{equation}
\beta(t) = \frac{y_{tgt}-y}{L}~(f_2 |\theta| + f_3 |\hat{\theta}|)
\end{equation}
Here, $\theta$ represents the follower's yaw angle about the $z$-axis, and $\hat{\theta}$ is its exponential moving average: $\hat{\theta}_{t+1} = \tfrac{1-\Delta t}{T_p} \hat{\theta}_t + \tfrac{\Delta t}{T_p} \theta$. The swimmers' $z$-positions remain fixed at $z_{tgt}$, as out-of-plane motion is not permitted. The controller-coefficients were selected to have a minimal impact on regular swimming kinematics, which allows for a direct comparison of the follower's efficiency to that of the leader:
\begin{eqnarray}
f_1 &=& 1 \\
f_2 &=& \max(0 , 50~\textrm{sign}(     \theta \cdot(y_{tgt}-y)) )  \\
f_3 &=& \max(0 , 20~\textrm{sign}(\hat{\theta}\cdot(y_{tgt}-y)) ) 
\end{eqnarray}





\newpage

\section*{Supporting Information - Supplementary Text, Figures, and Movies}
\setcounter{figure}{0}
\renewcommand{\figurename}{Figure S}

\paragraph*{Body-deformation during autonomous manoeuvres.}
The extent of body-bending that swimmers \Aeta and \Ady undergo when manoeuvring is compared quantitatively in Supplementary Fig.~S\ref{fig:contortionQuantitative}. A qualitative comparison was presented in Fig.~\ref{fig:contortion}.
\begin{figure}
    \centering
        \includegraphics{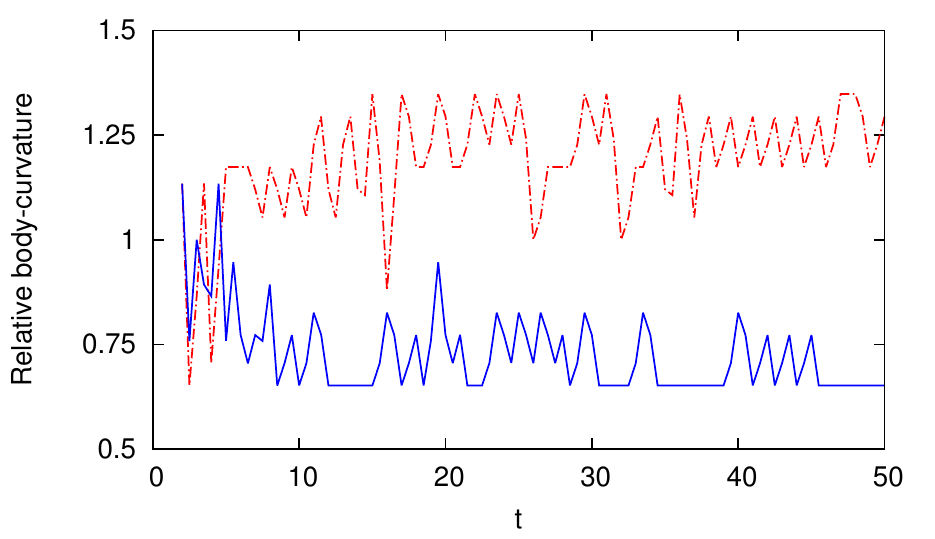}
        \caption{\textbf{Midline curvature.} Severity of body-deformation for the swimmers \Aeta (solid blue line) and \Ady (dash-dot red line), shown for 50 tail-beat periods starting from rest. The relative body-curvature is computed as $\Sigma_{i=1}^6\lvert \kappa_i \rvert$, normalized with the same metric for a solitary swimmer executing steady motion ($\kappa_i$ represents the curvature at 6 control points along a swimmer's body).}
        \label{fig:contortionQuantitative}
\end{figure}
We observe that the body-deformation of \Ady is noticeably higher than that of a steady swimmer (with relative curvature $1$), which implies a tendency to take aggressive turns. The deformation for swimmer \Aeta is markedly lower, which plays an instrumental role in reducing the power required for undulating the body against flow-induced forces.

\paragraph*{Comparison of four different swimmers.}
The performance metrics for four different swimmers are compared in Supplementary Fig.~S\ref{fig:compareSolo}.
\begin{figure*}
        \centering
        \begin{subfigure}[b]{\textwidth}
                \centering
                \includegraphics{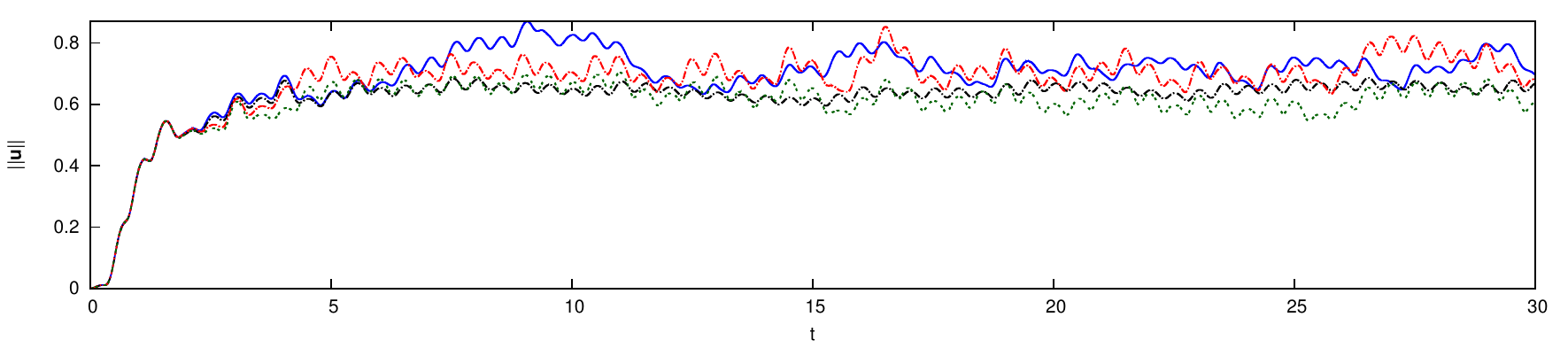}
                \subcaption{}
                \label{fig:speedComparison}
        \end{subfigure}
        \begin{subfigure}[b]{\textwidth}
                \centering
                \includegraphics{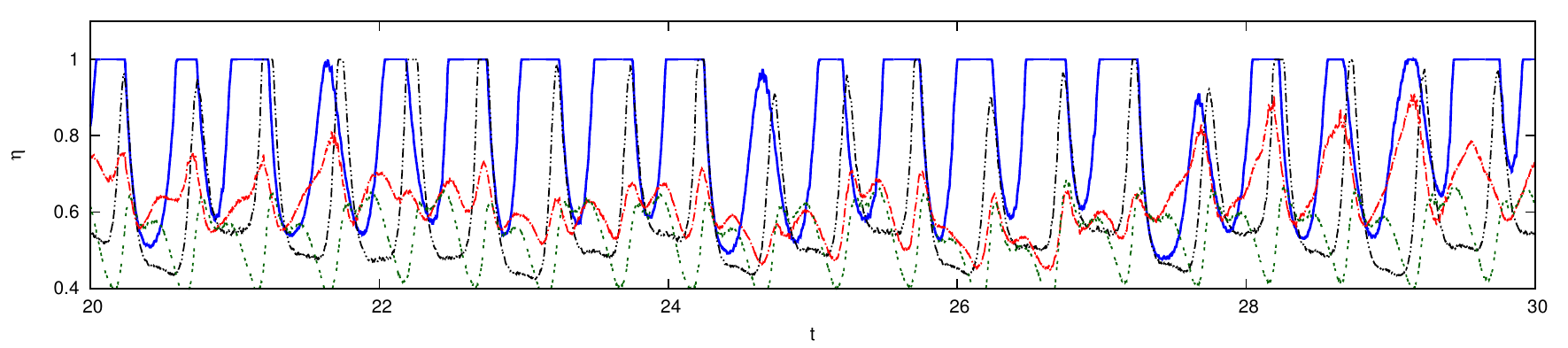}
                \subcaption{}
                \label{fig:etaComparison}
        \end{subfigure}
        \begin{subfigure}[b]{\textwidth}
                \centering
                \includegraphics{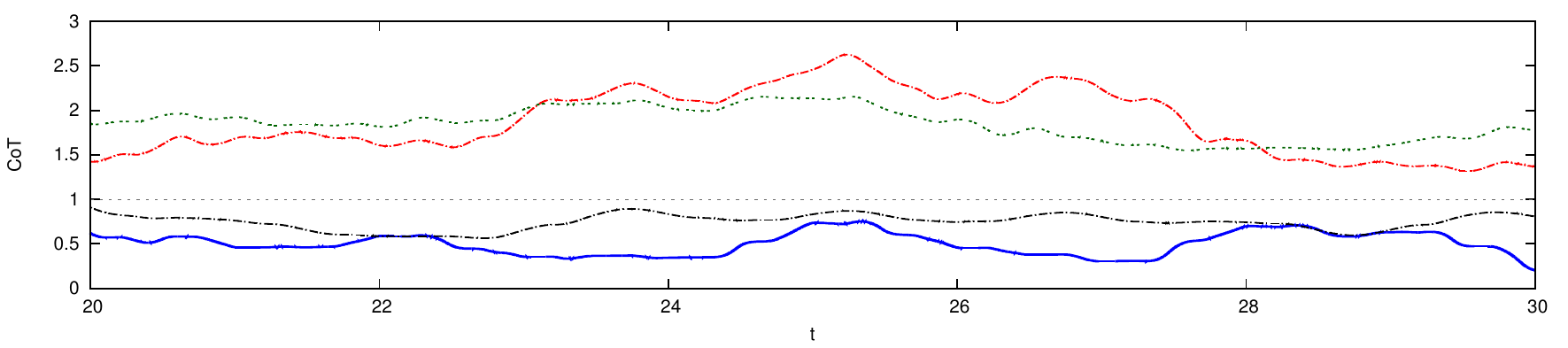}
                \subcaption{}
                \label{fig:cotComparison}
        \end{subfigure}
        \begin{subfigure}[b]{\textwidth}
                \centering
                \includegraphics{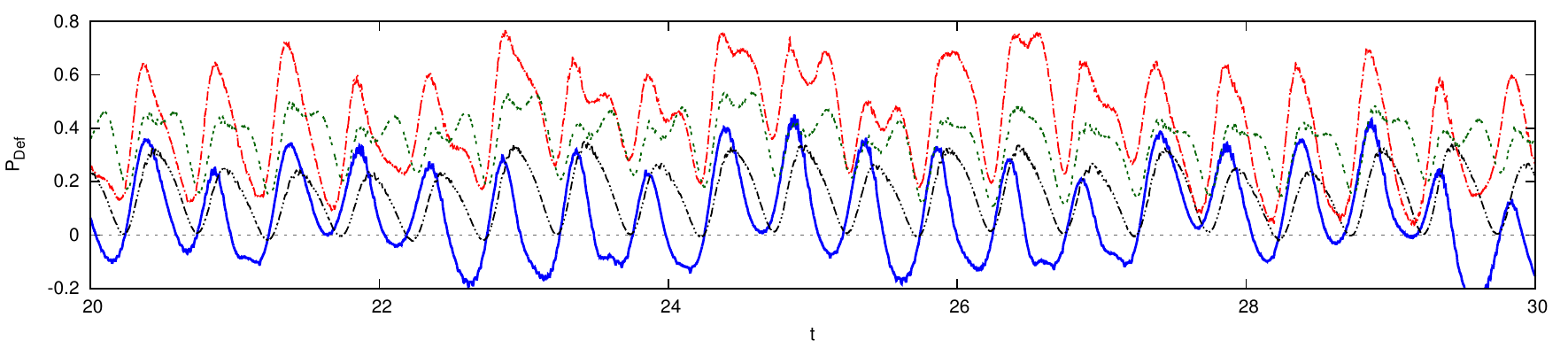}
                \subcaption{}
                \label{fig:pDefComparison}
        \end{subfigure}
         \begin{subfigure}[b]{\textwidth}
                \centering
                \includegraphics{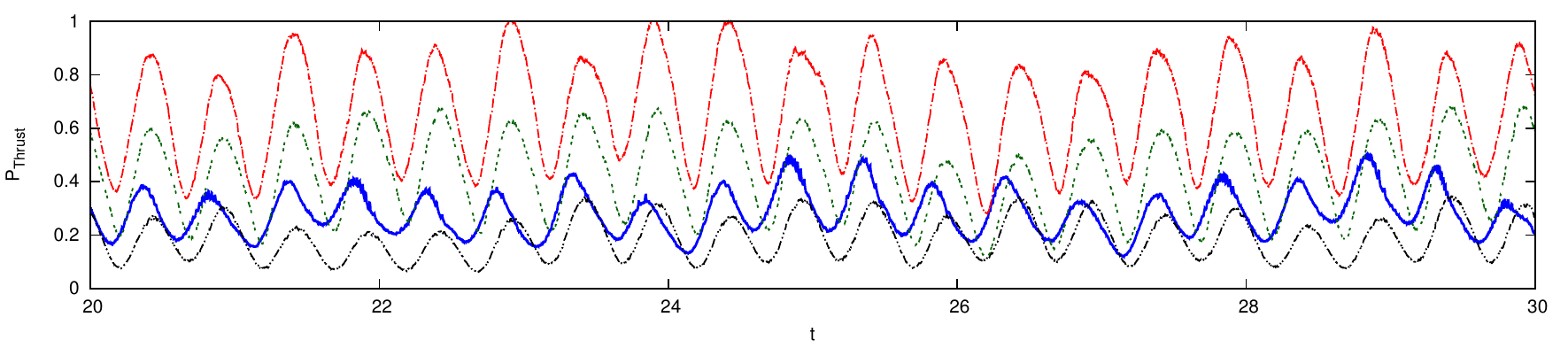}
                \subcaption{}
                \label{fig:pThrustComparison}
        \end{subfigure}
  \caption{\textbf{Performance metrics for four different swimmers.} Plots comparing (\subref{fig:speedComparison}) speed, (\subref{fig:etaComparison}) $\eta$, (\subref{fig:cotComparison}) CoT, (\subref{fig:pDefComparison}) deformation-power , and (\subref{fig:pThrustComparison}) thrust-power for four different swimmers. The solid blue line corresponds to swimmer \Aeta, the dash-double-dot black line to swimmer \Seta (a solitary swimmer executing actions identical to \Aeta), the dash-dot red line to swimmer \Ady, and the double-dot green line to swimmer \Sdy (a solitary swimmer executing actions identical to \Ady).}
\label{fig:compareSolo}
\end{figure*}
Interacting swimmer \Ady occasionally attains higher speed than \Aeta (Supplementary Fig.~S\ref{fig:speedComparison}), but at the cost of much higher energy expenditure (Supplementary Fig.~S\ref{fig:cotComparison} and Table~\ref{tab:compare}). 
\begin{table}[h]
\centering
\begin{tabular}{l|cccc}
	&\Aeta	&\Seta	&\Ady	&\Sdy \\
\hline
\hline
$\eta$	 	& 1.0	& 0.76 	& 0.77	& 0.66	\\
CoT      	& 1.0	& 1.56	& 3.96	& 3.86  \\
$P_{Def}$       & 1.0	& 1.41 	& 3.90	& 3.28	\\
$P_{Thrust}$	& 1.0	& 0.66 	& 2.33	& 1.48	\\
\end{tabular}
\caption{\textbf{Comparison of energetics metrics for the four swimmers.} Averaged values computed for the data shown in Supplementary Fig.~S\ref{fig:compareSolo}. All the values shown have been normalized with respect to the corresponding value for \Aeta.}
\label{tab:compare}
\end{table}
Moreover, the speeds of solitary swimmers \Seta and \Sdy are lower than those of either interacting swimmer (\Aeta and \Ady), which suggests that wake-interactions may benefit a follower regardless of the goal being pursued. In Supplementary Fig.~S\ref{fig:pDefComparison} $P_{Def}$ attains negative values only for \Aeta, which is indicative of maximum benefit extracted from flow-induced forces. Both \Ady and \Sdy are capable of generating significantly higher thrust-power than \Aeta, but suffer from larger deformation-power, and consequently, lower swimming-efficiency. Comparing the columns for \Aeta and \Seta in Table~S\ref{tab:compare}, we note that interacting with a preceding wake has a measurable impact on swimming-performance; \Aeta is approximately $32\%$ more efficient than \Seta, spends $36\%$ less energy per unit distance travelled, requires $29\%$ less power for body-undulations, and generates $52\%$ higher thrust-power. Wake-interactions yield energetics benefits even for the swimmer actively minimizing lateral displacement from the leader, primarily by increasing thrust-power, as can be surmised by comparing the data for \Ady and \Sdy in Supplementary Table~\ref{tab:compare}.

\begin{figure}
    \centering
	\includegraphics{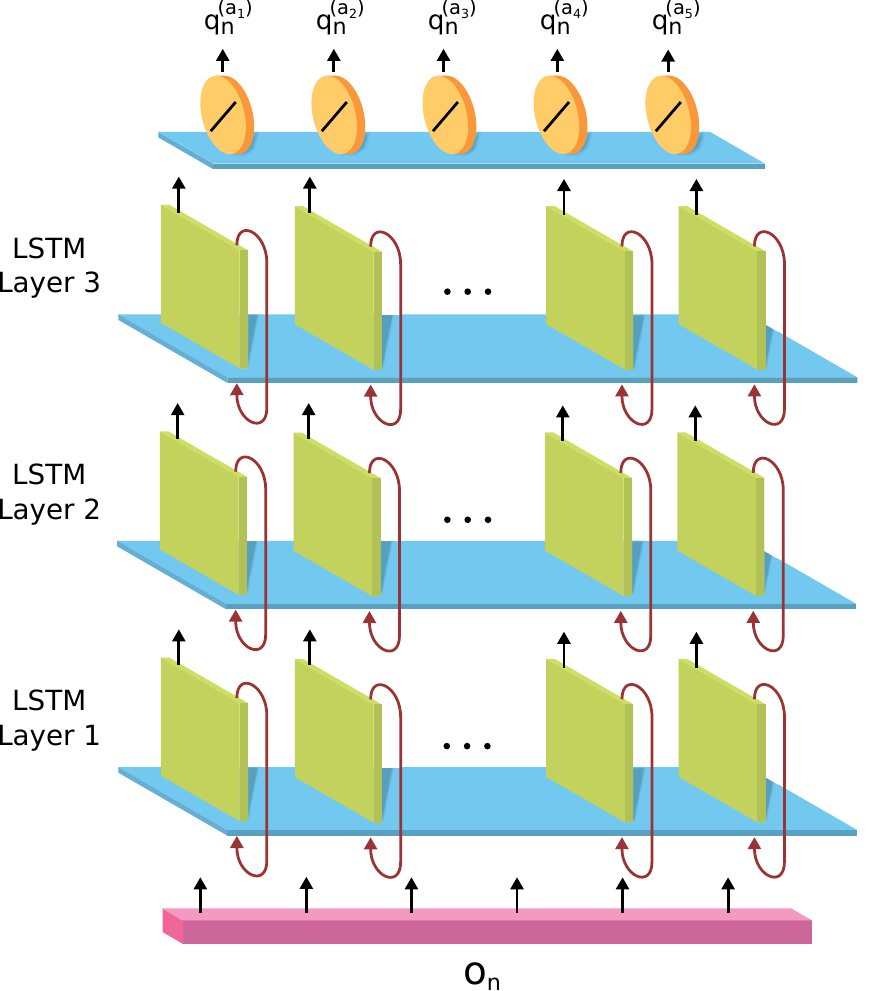}
  	\caption{\textbf{Schematic of the Recurrent Neural Network (RNN).} The RNN used in this study is composed of 3 LSTM layers, consisting of 24 cells (green blocks) each. The input layer (pink block) of the network comprises the 6 observed-state variables. The black arrows between different layers indicate all-to-all connections. The purple arrows indicate recurrent connections within each LSTM layer. The last layer consists of 5 output neurons (orange) with linear activation.}
	\label{fig:NNstructure}
\end{figure}

\paragraph*{Uncovering underlying time-dependencies.}
While it is relatively straightforward to maintain a particular tandem formation via feedback control (when the follower strays too far to one side, a feedback controller can relay instructions to veer in the opposite direction), the same is not true for maximizing swimming-efficiency. It is difficult to formulate a simple set of a-priori rules for maximizing efficiency, especially in dynamically evolving conditions. This happens because: 1) the swimmer perceives only a limited representation of its environment (Fig.~\ref{fig:states}); and 2) there may be measurable delay between an action and its impact on the reward received over the long term. These traits make deep RL ideal for determining the optimal policy when maximizing swimming-efficiency, especially when augmented with recurrent neural networks (Supplementary Fig.~S\ref{fig:NNstructure}). These network architectures are adept at discovering and exploiting long-term time-dependencies.

\paragraph*{Flow-interactions at the instant of minimum swimming-efficiency.}
The instant when swimmer \Aeta attains the lowest efficiency during each half-period ($\eta_{min}(D)$ in Fig.~\ref{fig:etaPlot}) is examined in Supplementary Fig.~S\ref{fig:fluidsPointEtaMinD}.
\begin{figure*}
        \centering
        \begin{subfigure}[b]{0.59\textwidth}
                \centering
                \includegraphics[width=1.0\textwidth]{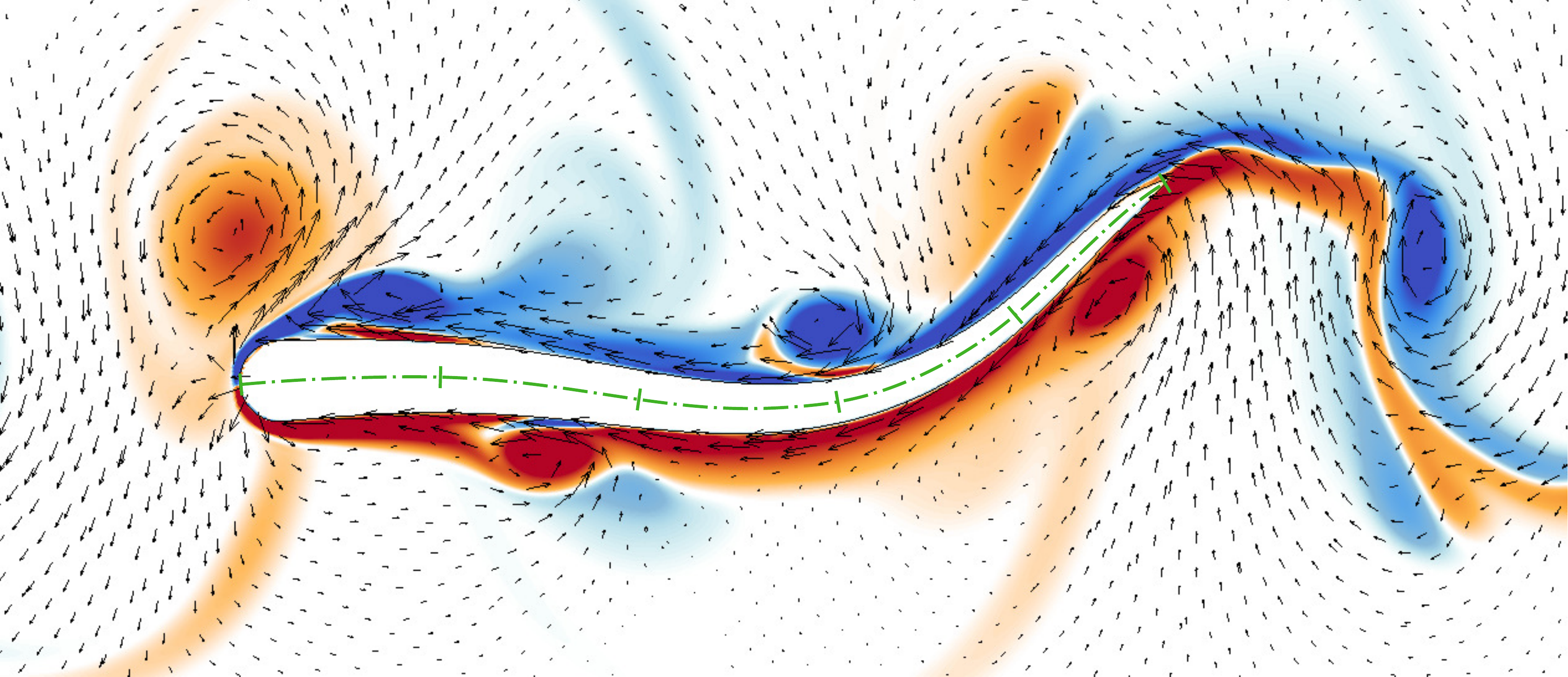}
                \includegraphics[width=1.0\textwidth]{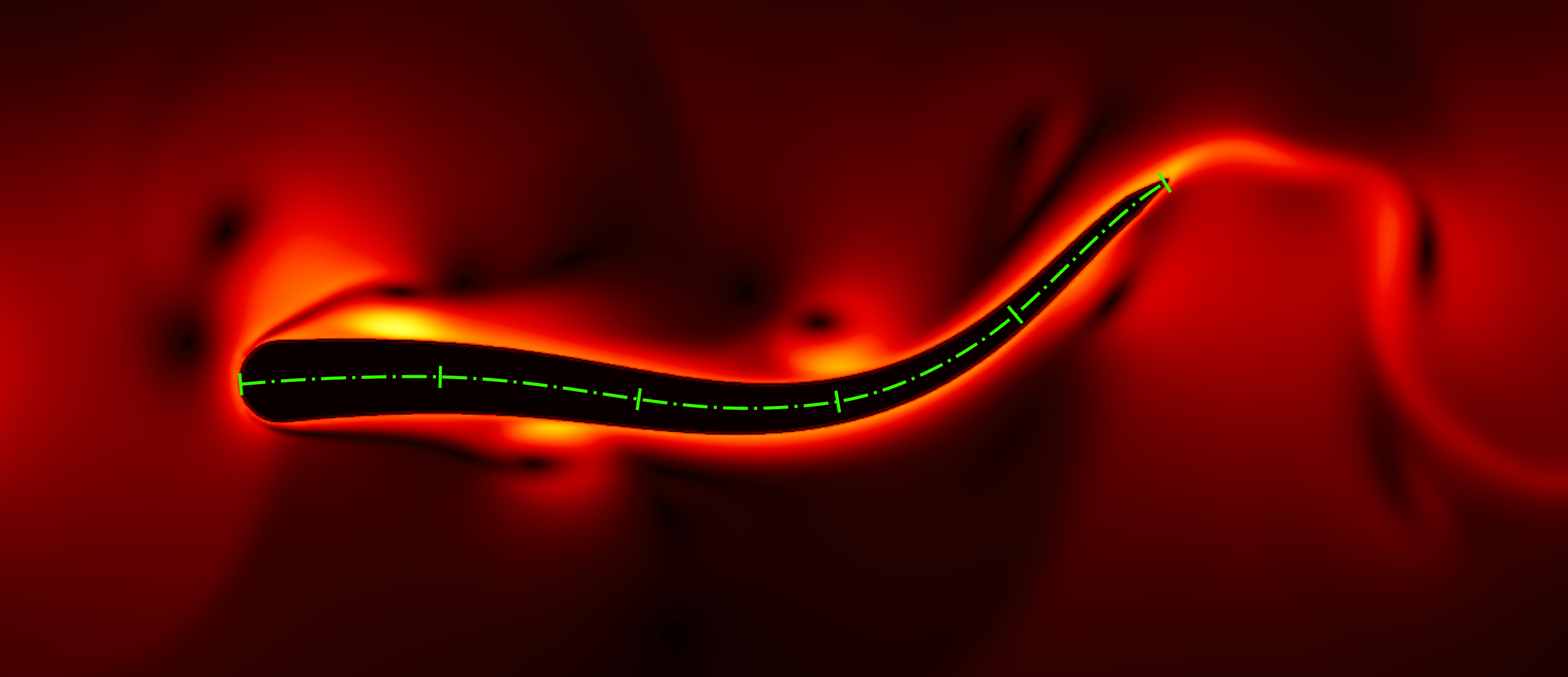}
                \subcaption{}
                \label{fig:vortVelEtaMinD}
        \end{subfigure}
        \begin{subfigure}[b]{0.39\textwidth}
                \centering
                \includegraphics{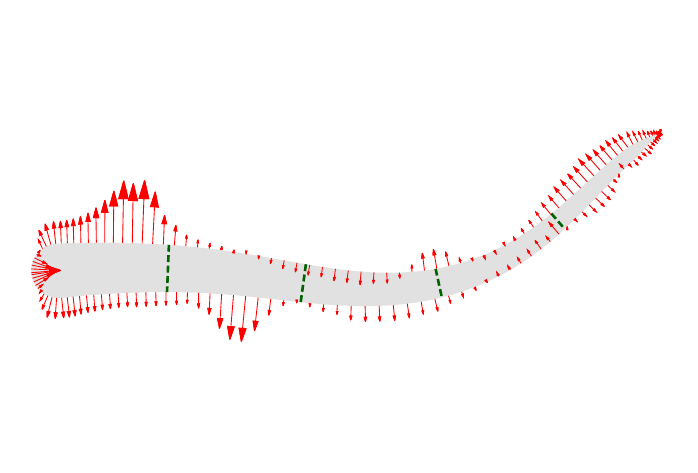}
                \includegraphics{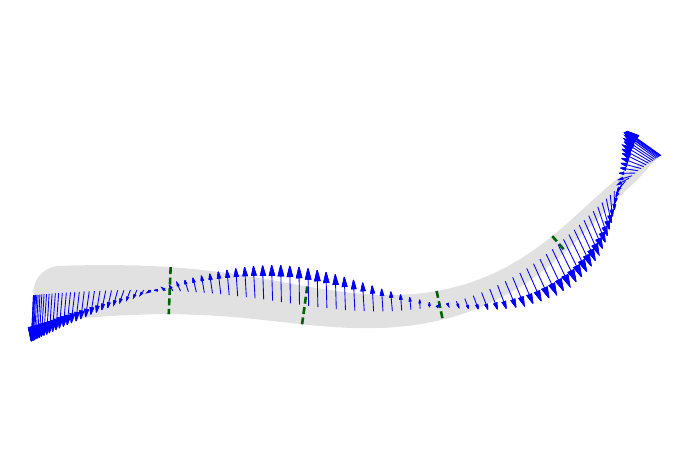}
                \subcaption{}
                \label{fig:kinematicsEtaMinD}
        \end{subfigure} \\
        \begin{subfigure}[b]{0.49\textwidth}
                \centering
                \includegraphics{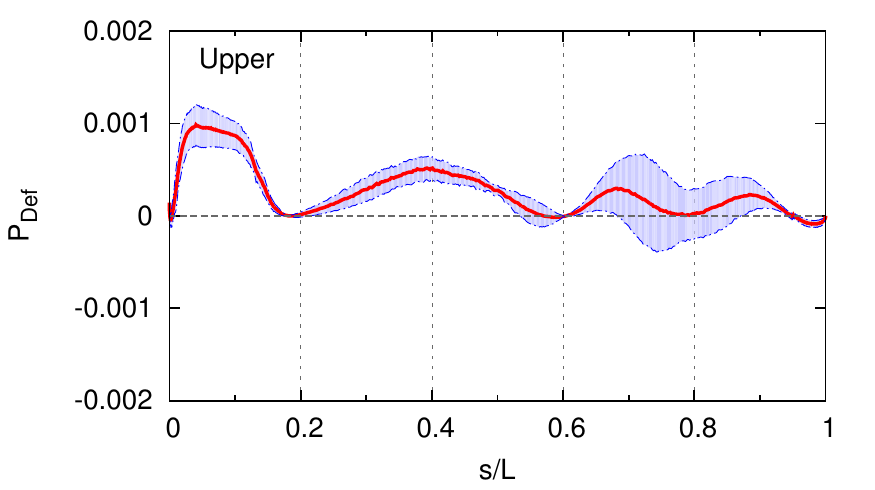}
                \subcaption{}
                \label{fig:pDefUEtaMinD}
        \end{subfigure}
        \begin{subfigure}[b]{0.49\textwidth}
                \centering
                \includegraphics{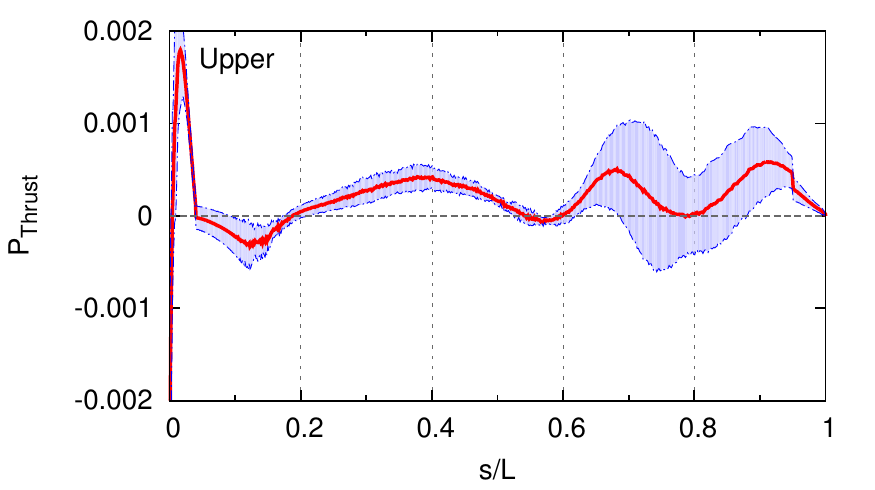}
                \subcaption{}
                \label{fig:pThrustUEtaMinD}
        \end{subfigure} \\
        \begin{subfigure}[b]{0.49\textwidth}
                \centering
                \includegraphics{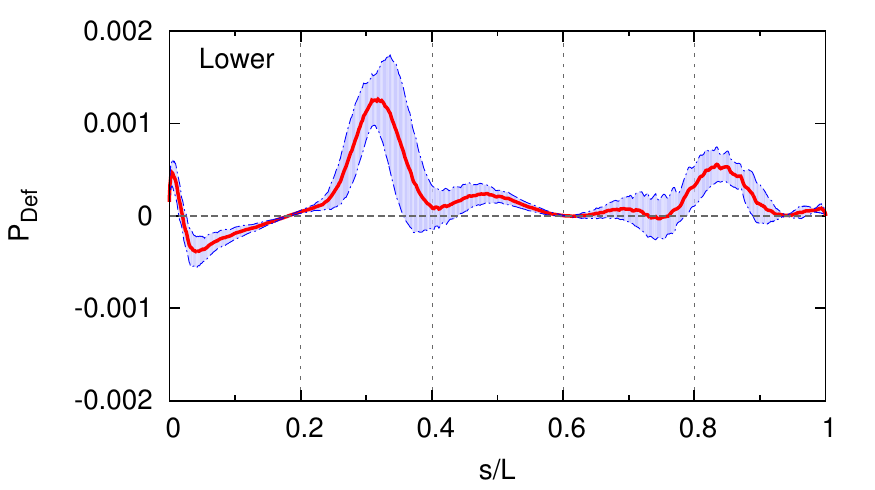}
                \subcaption{}
                \label{fig:pDefLEtaMinD}
        \end{subfigure}
        \begin{subfigure}[b]{0.49\textwidth}
                \centering
                \includegraphics{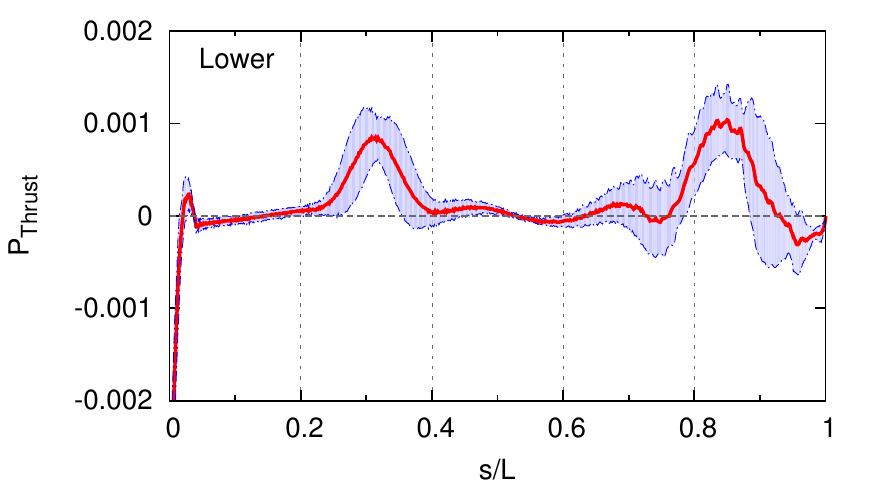}
                \subcaption{}
                \label{fig:pThrustLEtaMinD}
        \end{subfigure}
  \caption{\textbf{Flow-field and flow-induced forces for \Aeta, corresponding to minimum efficiency.} (\subref{fig:vortVelEtaMinD}) Vorticity field with the velocity vectors shown (top), and velocity magnitude (bottom) at $t=26.87$ (point $\eta_{min}(D)$ in Fig.~\ref{fig:rEtaData}). (\subref{fig:kinematicsEtaMinD}) Flow-induced force-vectors (top) and body-deformation velocity (bottom) at this instance. (\subref{fig:pDefUEtaMinD},\subref{fig:pThrustUEtaMinD}) Deformation-power and thrust-power acting on the upper (right lateral) surface of follower. The red line indicates the average over 10 different snapshots ranging from $t=30.87$ to $t=39.87$. The envelope denotes the standard deviation among the 10 snapshots. (\subref{fig:pDefLEtaMinD},\subref{fig:pThrustLEtaMinD}) Deformation-power and thrust-power on the lower (left lateral) surface of the fish.}
\label{fig:fluidsPointEtaMinD}
\end{figure*}
The mean $P_{Def}$ curve is mostly positive on both the lower and upper surfaces, with large positive peaks generated by interaction with the wake- and lifted-vortices. This increase in effort is not offset sufficiently by an increase in $P_{Thrust}$, resulting in low swimming-efficiency. Compared to the instance of maximum efficiency (Fig.~\ref{fig:fluidsEtaMaxA}), increased effort is required in the head region, along with an increase in thrust-production by the tail section $s>0.7L$. 

\paragraph*{Slight deviations impact performance.}
To examine the impact of small deviations in \Aeta's trajectory on its performance, we compare two different time-instances (at the same tail-beat stage) in Supplementary Fig.~S\ref{fig:examinePointE}.
\begin{figure*}
        \centering
        \begin{subfigure}[b]{0.49\textwidth}
                \centering
                \includegraphics[width=1.0\textwidth]{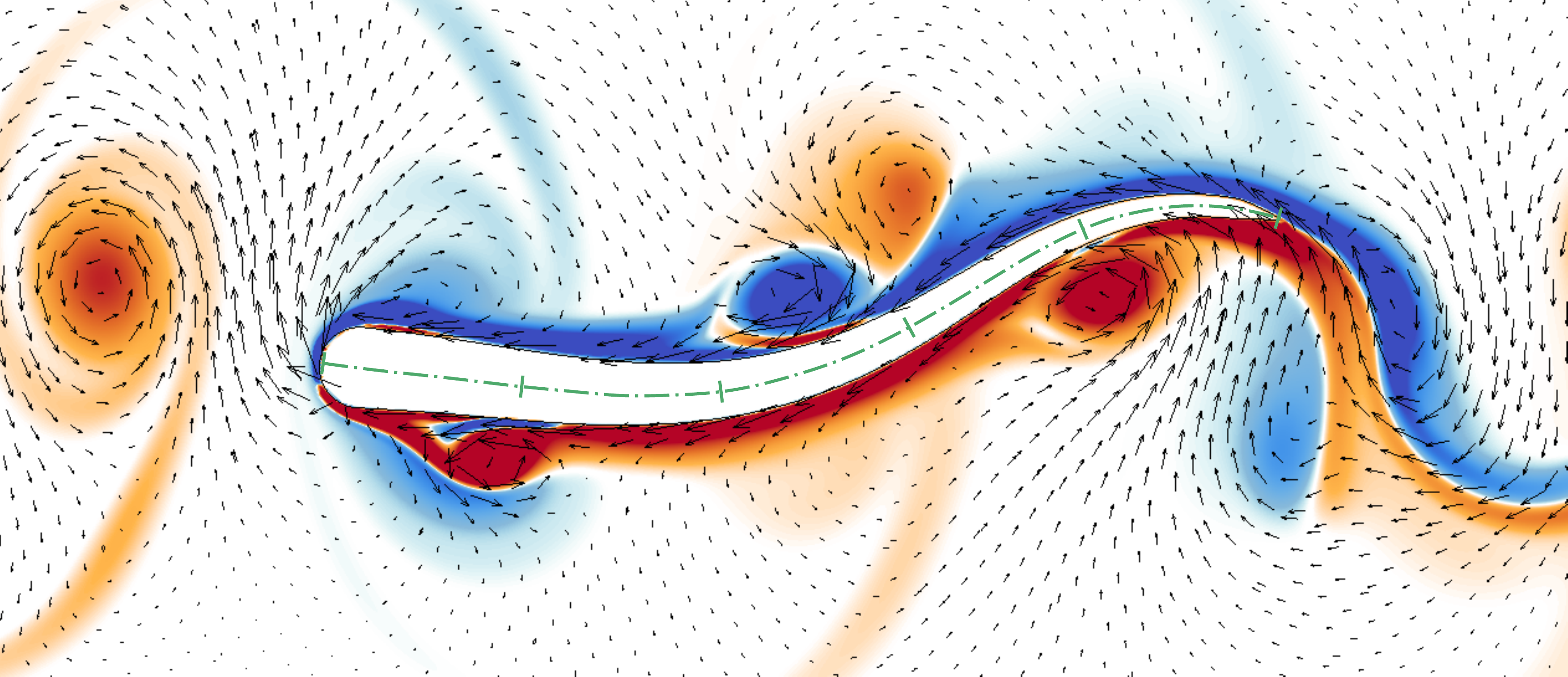}
                \includegraphics[width=1.0\textwidth]{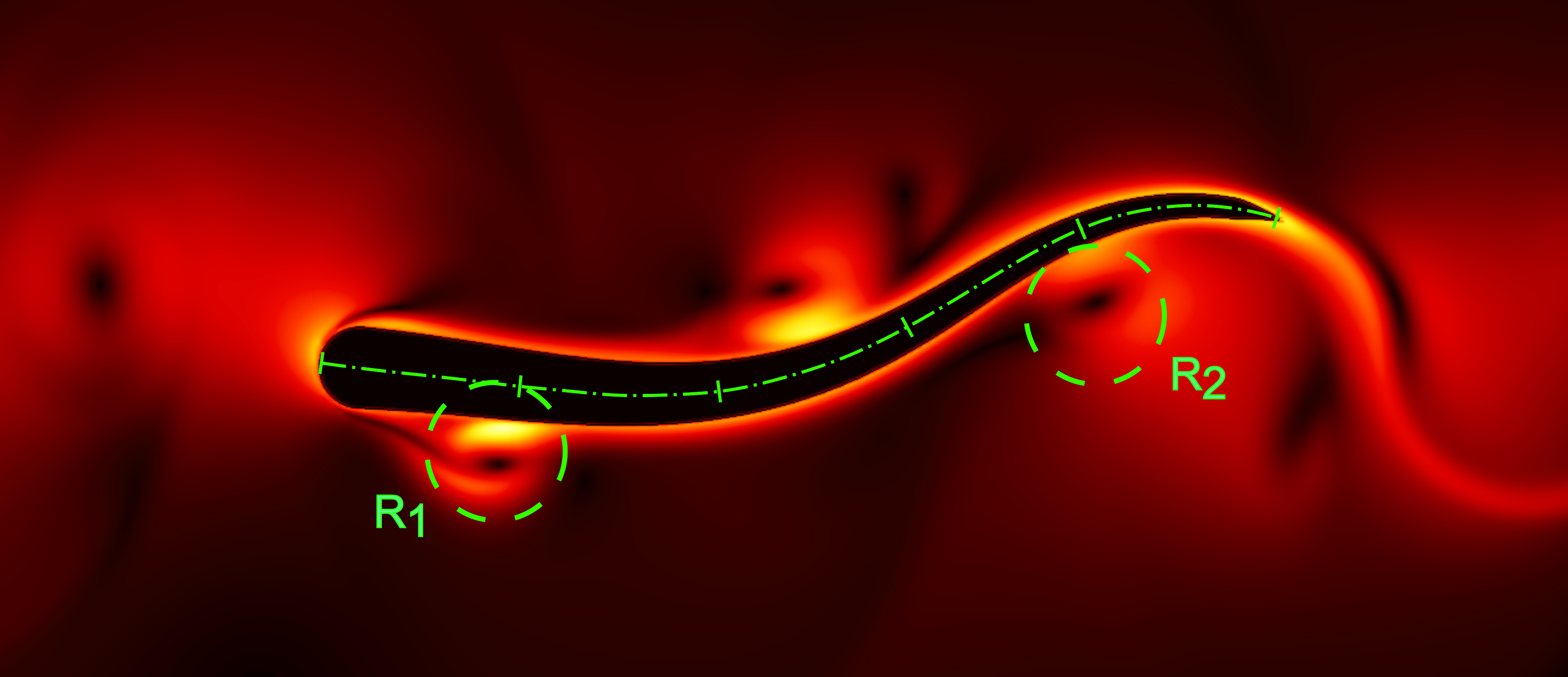}
                \subcaption{}
                \label{fig:vortVelEtaMaxC}
        \end{subfigure}
        \begin{subfigure}[b]{0.49\textwidth}
                \centering
                \includegraphics[width=1.0\textwidth]{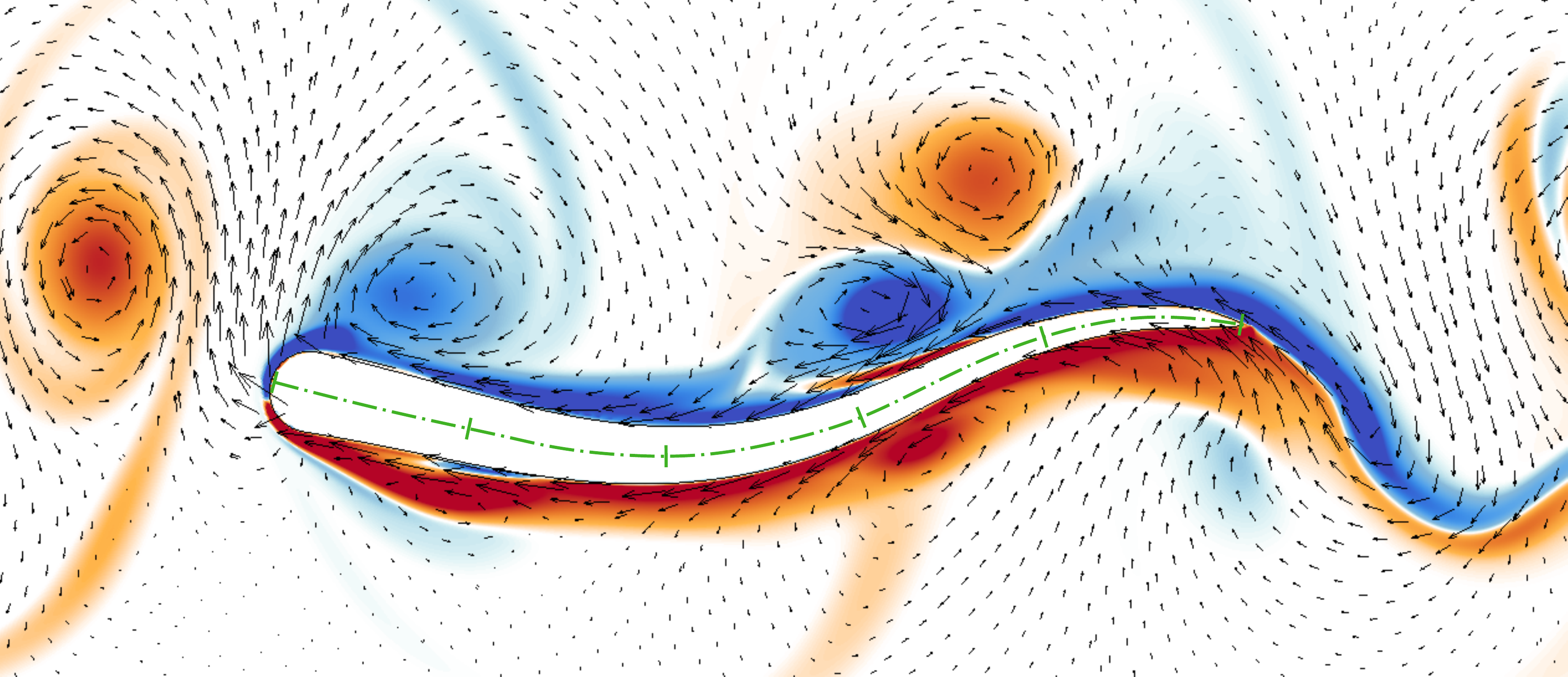}
                \includegraphics[width=1.0\textwidth]{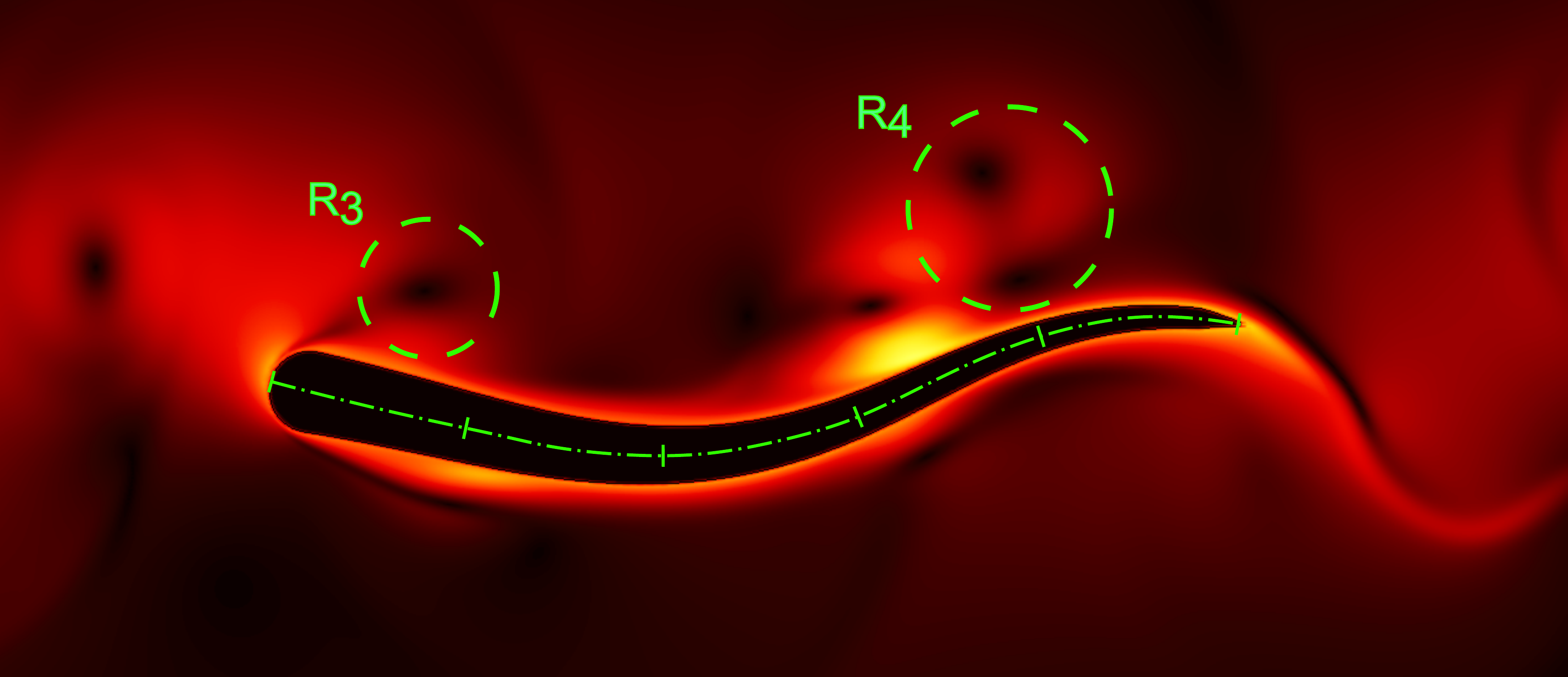}
                \subcaption{}
                \label{fig:vortVelEtaPointE}
        \end{subfigure}\\
        \begin{subfigure}[b]{0.49\textwidth}
                \centering
                \includegraphics{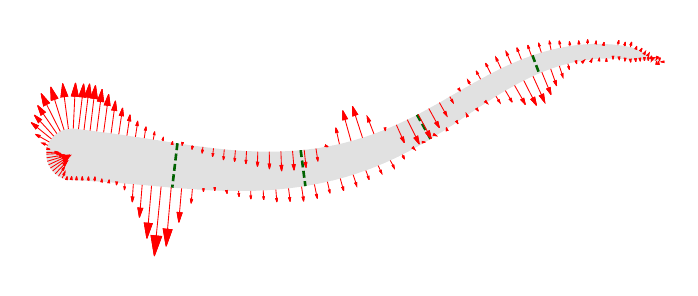}
                \includegraphics{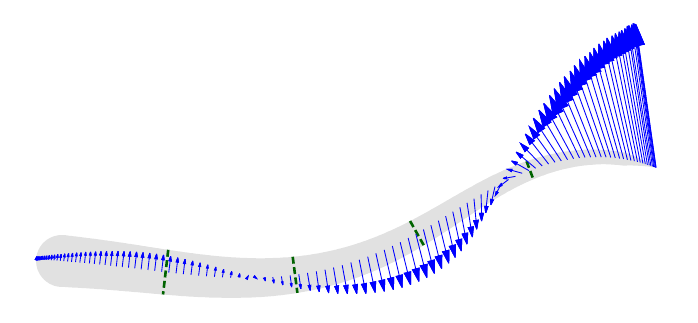}
                \subcaption{}
                \label{fig:forceEtaMaxC}
        \end{subfigure}
        \begin{subfigure}[b]{0.49\textwidth}
                \centering
                \includegraphics{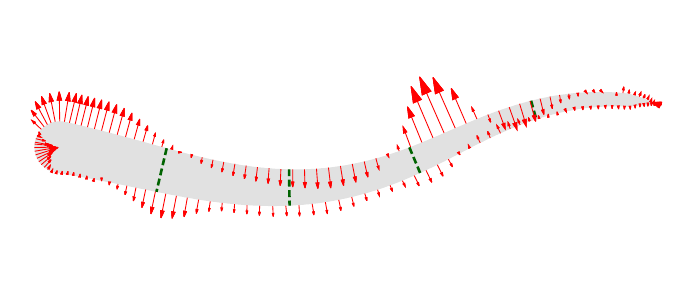}
                \includegraphics{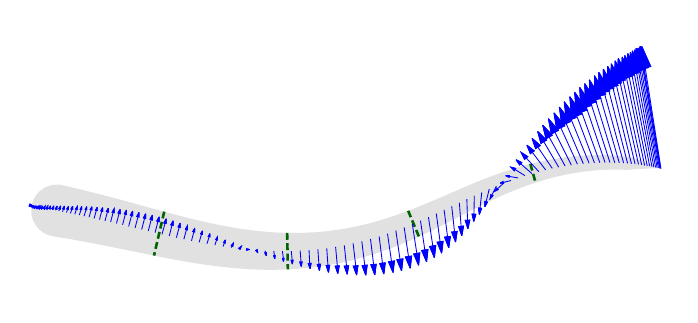}
                \subcaption{}
                \label{fig:forceEtaPointE}
        \end{subfigure}\\
        \begin{subfigure}[b]{0.49\textwidth}
                \centering
                \includegraphics{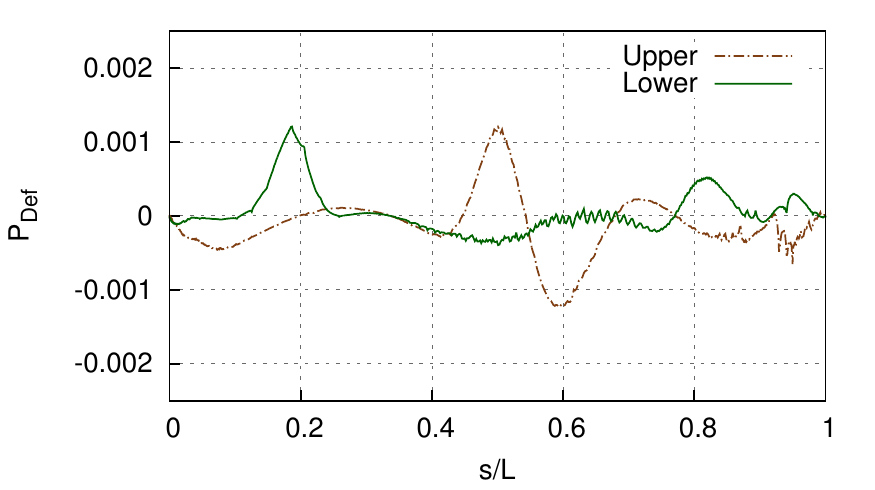}
                \subcaption{}
                \label{fig:pDefEtaMaxC}
        \end{subfigure}
        \begin{subfigure}[b]{0.49\textwidth}
                \centering
                \includegraphics{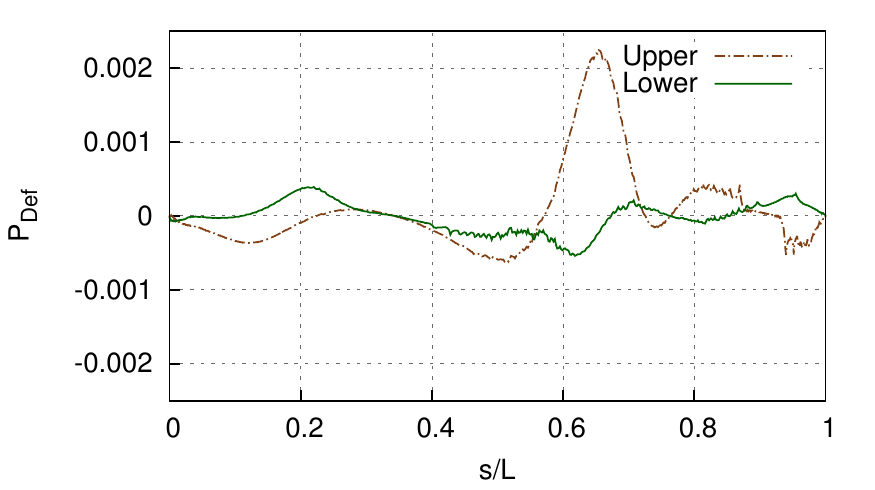}
                \subcaption{}
                \label{fig:pDefEtaPointE}
        \end{subfigure}
  \caption{\textbf{Deviations impact performance.} Comparison of two instances when a maximum in the swimming-efficiency is expected. The deformed shape and deformation-velocity for the two instances are similar, but differences in the flow-field influence efficiency. Panels on the left hand side of the page show data for \Aeta at $t\approx33.7$ ($\eta=1$), whereas those on the right hand side correspond to $t\approx27.7$ ($\eta=0.86$). (\subref{fig:vortVelEtaMaxC}, \subref{fig:vortVelEtaPointE}) Vorticity, velocity vectors, and velocity magnitude at the two time instances. A slight deviation in the follower's approach to the wake causes a noticeable change in the surrounding vortices, as well as in the velocity induced near the surface. The regions highlighting differences have been marked as $R_1$, $R_2$, $R_3$, and $R_4$. (\subref{fig:forceEtaMaxC}, \subref{fig:forceEtaPointE}) A comparison of the surface force-vectors and body-deformation velocity. (\subref{fig:pDefEtaMaxC},\subref{fig:pDefEtaPointE}) There are notable differences in the distribution of $P_{Def}$ on the upper and lower surfaces.}
\label{fig:examinePointE}
\end{figure*}
At $t\approx26.5$, \Aeta deviates slightly to the left of its steady trajectory (\MovieFour), which throws it out of synchronization with the oncoming wake-vortices. The resulting reduction in efficiency at $t\approx27.5$ indicates that even slight deviations are capable of impacting performance, and that there may be a measurable delay between actions and consequences. However, the smart-swimmer autonomously corrects for such deviations, and is able to quickly recover its optimal behaviour. 

\paragraph*{Correlation with the flow-field}
The correlation-coefficient curve shown in Fig.~\ref{fig:histogram}, and the correlation map shown in Supplementary Fig.~S\ref{fig:correlation3D}, were computed as follows:
\begin{equation}
\rho(\bm{u},\bm{u}_{\textrm{head}}) = \dfrac{\textrm{cov}\left(\bm{u}(x,y),\bm{u}_{\textrm head}\right)}{\sigma_{\bm{u}(x,y)} ~ \sigma_{\bm{u}_{\textrm{head}}}} = \dfrac{\sum_{t} \bm{u}(x,y,t) \cdot \bm{u}_{\textrm{head}}(t)} {\sqrt{\sum_{t} \| \bm{u}(x,y,t) \|^2} \sqrt{\sum_{t} \| \bm{u}_{\textrm{head}}(t)\|^2 } }
\label{eq:corrCoeff}
\end{equation}
\begin{figure}
    \centering
        \includegraphics[width=0.8\textwidth]{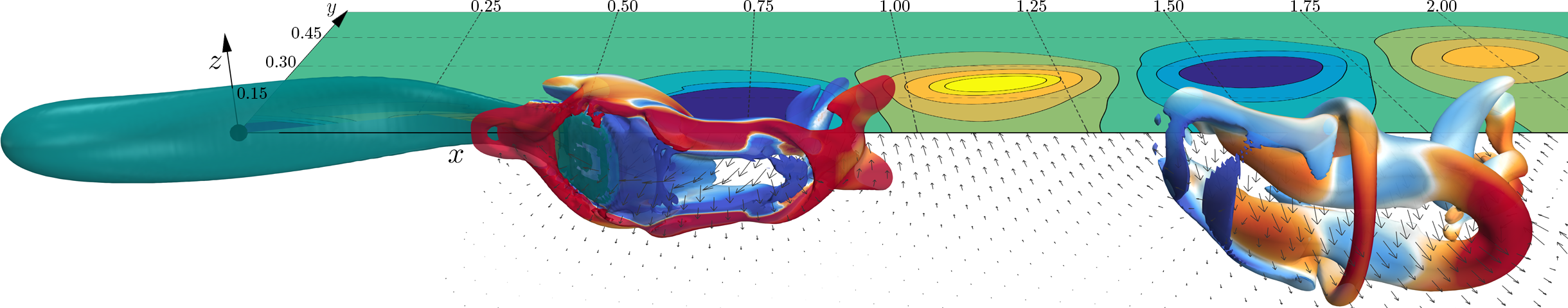}
        \caption{\textbf{Correlation map.} The horizontal plane on the right side of the swimmer depicts the correlation-coefficient described by Equation~\ref{eq:corrCoeff}. Areas of high correlation are denoted as yellow regions, whereas those of low correlation are shown in blue. The vortex rings shed are shown on the swimmer's left side, along with the velocity vectors on the left horizontal plane.}
        \label{fig:correlation3D}
\end{figure}
Here, $\bm{u}(x,y,t)$ was recorded in the wake of a solitary swimmer, whereas $\bm{u}_{\textrm{head}}(t)$ was recorded at the swimmer's head. Maxima in $\rho(\bm{u},\bm{u}_{\textrm{head}})$ provide an estimate for the coordinates where a follower's head-movements would exhibit long-term synchronization with an undisturbed wake.

\paragraph*{Limiting the exploration space.}
During training, the range of values that a smart-follower's states can take are constrained, as mentioned previously. This prevents excessive exploration of regions that involve no wake-interactions, and helps to minimize the computational cost of training-simulations. The limits of the bounding box (shown in Supplementary Fig.~S\ref{fig:reward}) are kept sufficiently large to provide the follower ample room to swim clear of the unsteady wake, if it determines that interacting with the wake is unfavourable.
\begin{figure}
    \centering
                \includegraphics[width=8.9cm]{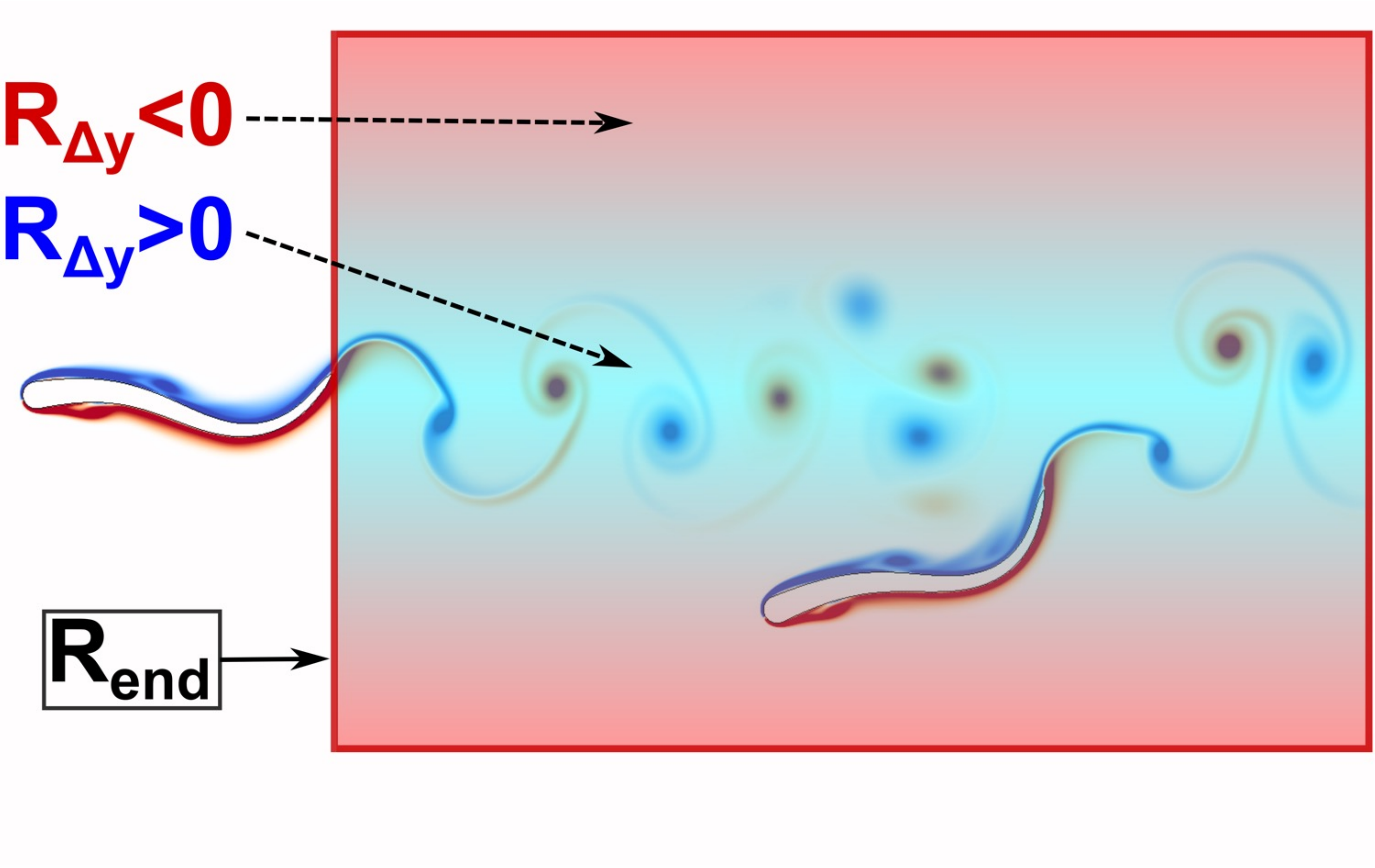}
        \caption{\textbf{Reward for \Ady.} Visual representation of reward assigned to smart-swimmer \Ady, whose goal is to minimize its lateral displacement from the leader.}
        \label{fig:reward}
\end{figure}

\paragraph*{Power distribution in the presence/absence of a preceding wake.}
To determine the extent to which wake-induced interactions alter the distribution of $P_{Def}$ and $P_{Thrust}$, both of which influence overall swimming-efficiency, we compare these quantities for \Aeta and \Seta in Supplementary Fig.~S\ref{fig:periodPowerDistribEta}.
\begin{figure*}
        \centering
        \begin{subfigure}[b]{8.9cm}
                \centering
                		\includegraphics{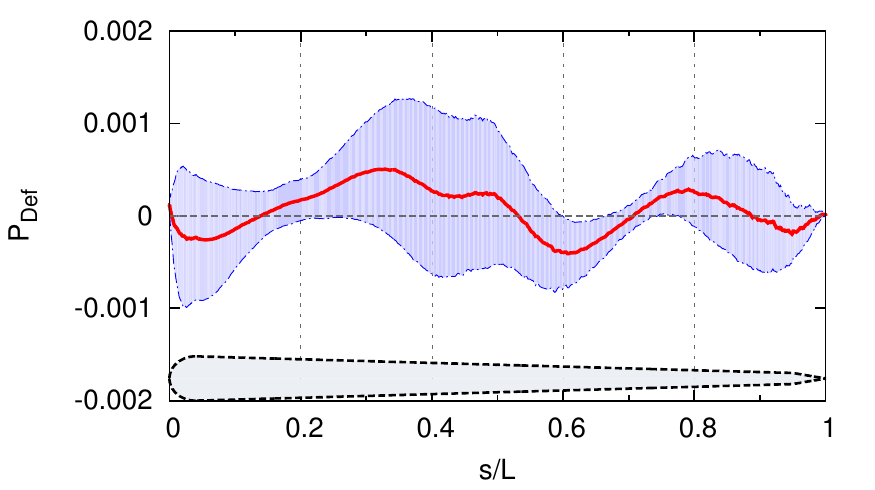}
                		\subcaption{}
                		\label{fig:pDefEnvelopeAeta}
        \end{subfigure}
         \begin{subfigure}[b]{8.9cm}
                \centering
                		\includegraphics{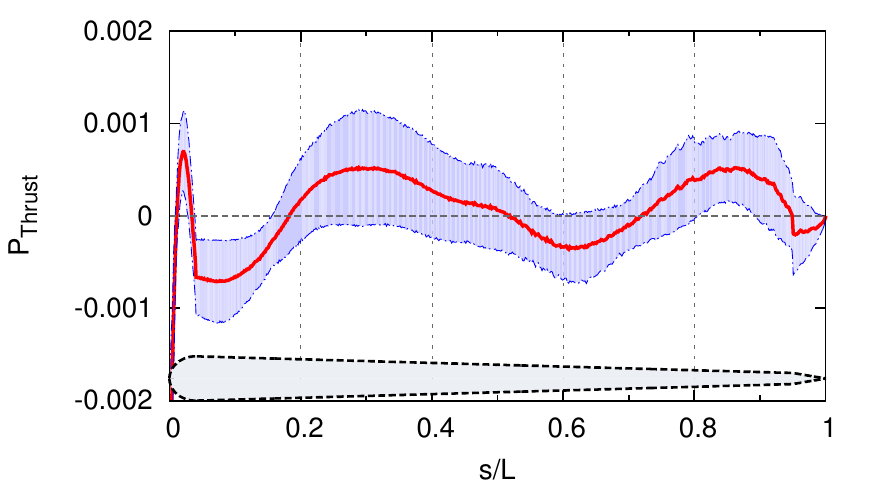}
                \subcaption{}
                \label{fig:pThrustEnvelopeAeta}
        \end{subfigure}
        \begin{subfigure}[b]{8.9cm}
                \centering
                		\includegraphics{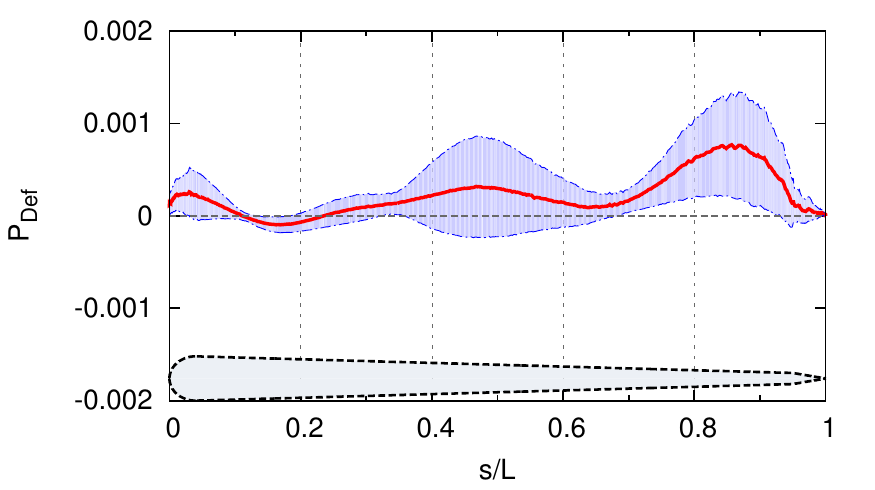}
                \subcaption{}
                \label{fig:pDefEnvelopeSeta}
        \end{subfigure}
        \begin{subfigure}[b]{8.9cm}
                \centering
                \includegraphics{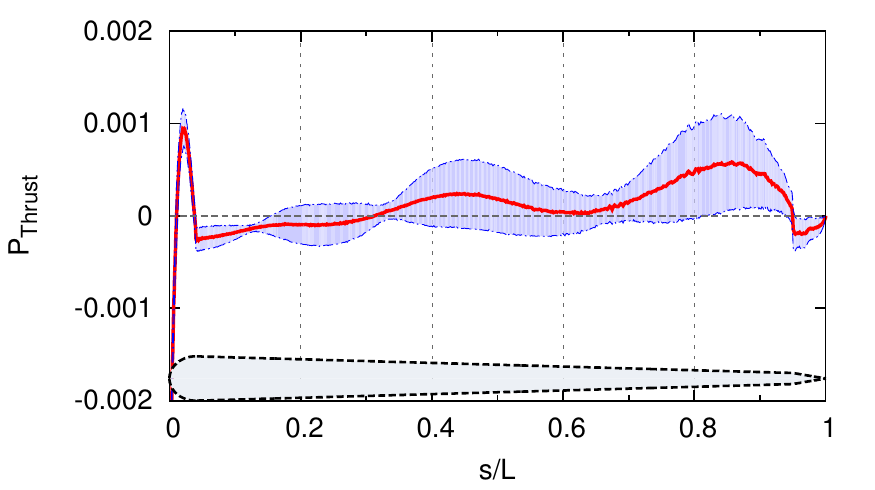}
                \subcaption{}
                \label{fig:pThrustEnvelopeSeta}
        \end{subfigure}
  		\caption{\textbf{Power distribution.} Deformation-power and thrust-power distribution along the body of  (\subref{fig:pDefEnvelopeAeta},\subref{fig:pThrustEnvelopeAeta}) swimmer \Aeta, and (\subref{fig:pDefEnvelopeSeta},\subref{fig:pThrustEnvelopeSeta}) swimmer \Seta. The solid red line indicates the average over a single tail-beat period (from $t=26$ to $t=27$), whereas the envelope denotes the standard-deviation. The silhouettes at the bottom of each panel represent the fish body.}
\label{fig:periodPowerDistribEta}
\end{figure*}
A similar comparison for \Ady and \Sdy is shown in Supplementary Fig.~S\ref{fig:periodPowerDistribYdisp}.
\begin{figure*}
        \centering
         \begin{subfigure}[b]{8.9cm}
                \centering
                \includegraphics{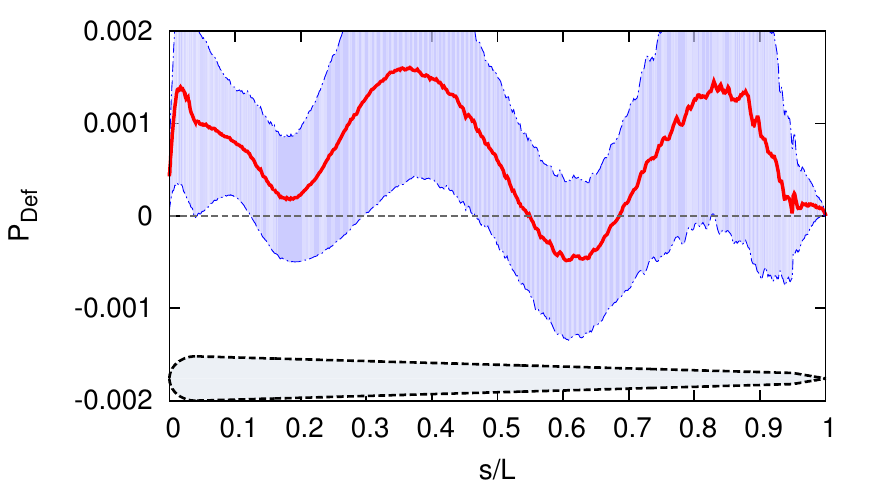}
                \subcaption{}
                \label{fig:pDefEnvelopeAdy}
        \end{subfigure}
         \begin{subfigure}[b]{8.9cm}
                \centering
                \includegraphics{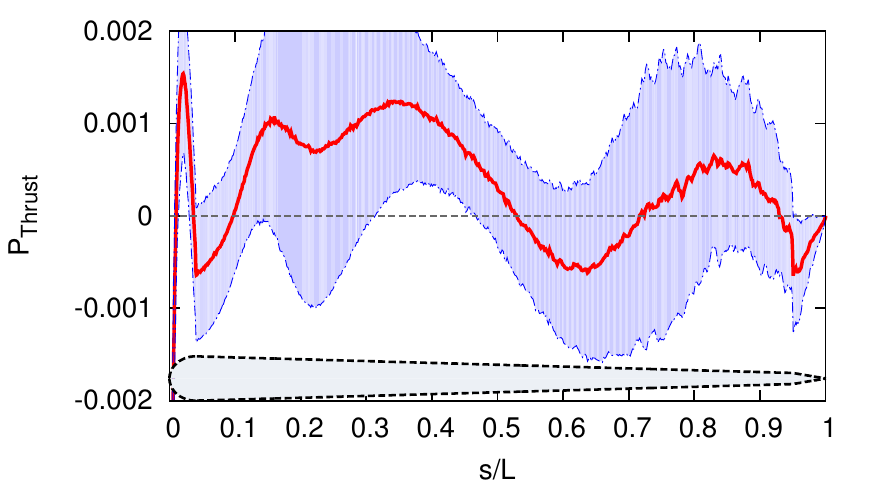}
                \subcaption{}
                \label{fig:pThrustEnvelopeAdy}
        \end{subfigure}
         \begin{subfigure}[b]{8.9cm}
                \centering
               	\includegraphics{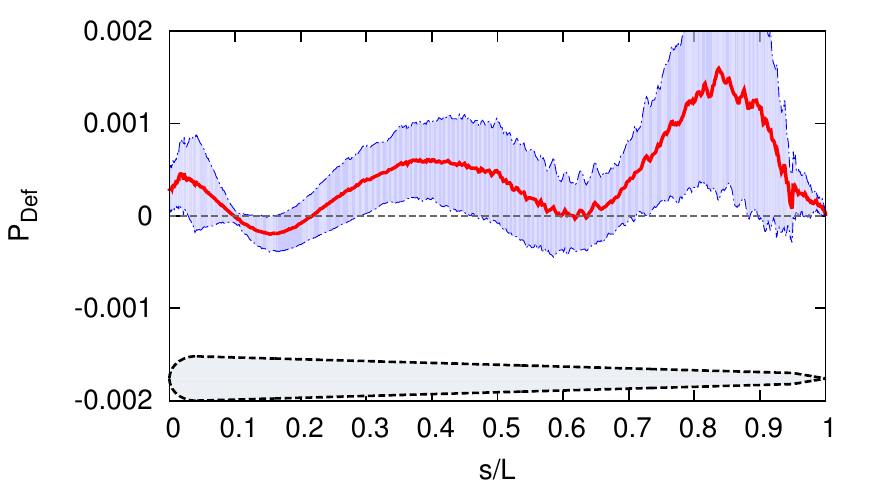}
                \subcaption{}
                \label{fig:pDefEnvelopeSdy}
        \end{subfigure}
         \begin{subfigure}[b]{8.9cm}
                \centering

             	\includegraphics{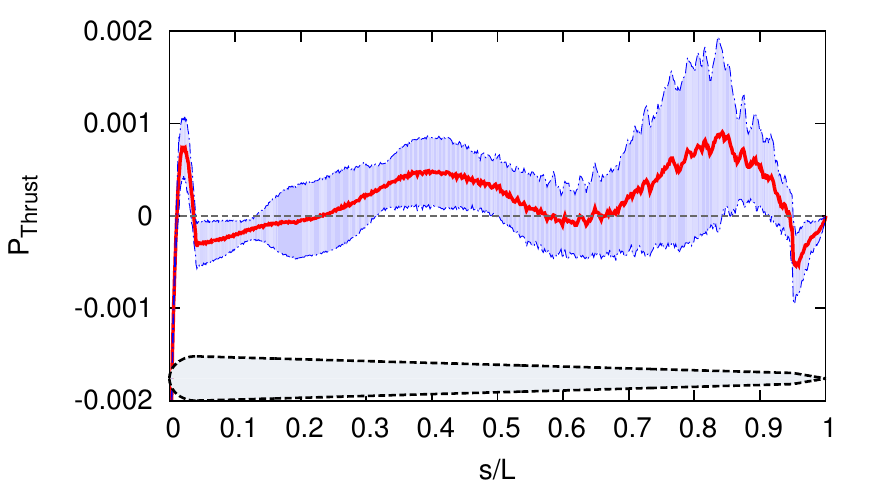}
                \subcaption{}
                \label{fig:pThrustEnvelopeSdy}
        \end{subfigure}
  \caption{\textbf{Power distribution}. Deformation-power and thrust-power distribution along the body of (\subref{fig:pDefEnvelopeAdy}, \subref{fig:pThrustEnvelopeAdy}) swimmer \Ady, and (\subref{fig:pDefEnvelopeSdy}, \subref{fig:pThrustEnvelopeSdy}) swimmer \Sdy. The solid red line indicates the average over a single tail-beat period (from $t=26$ to $t=27$), whereas the envelope denotes the standard-deviation. The silhouettes at the bottom of each panel represent the fish body.}
\label{fig:periodPowerDistribYdisp}
\end{figure*}
For \Aeta, a greater variation in $P_{Def}$ and $P_{Thrust}$ is observed (broad envelopes in Supplementary Figs.~S\ref{fig:pDefEnvelopeAeta} and~S\ref{fig:pThrustEnvelopeAeta}), compared to the solitary swimmer \Seta (Supplementary Figs.~S\ref{fig:pDefEnvelopeSeta} and~S\ref{fig:pThrustEnvelopeSeta}). This is caused by \Aeta's interactions with the unsteady wake, which is absent for \Seta.  The average $P_{Def}$ for \Aeta shows distinct negative troughs near the head ($s/L<0.2$, Supplementary Fig.~S\ref{fig:pDefEnvelopeAeta}) and at $s/L=0.6$. A lack of similar troughs for \Seta (Supplementary Fig.~S\ref{fig:pDefEnvelopeSeta}) implies that these benefits originate exclusively from wake-induced interactions. There is no apparent difference in drag for both \Aeta and \Seta in the pressure-dominated region close to the head ($s\approx0$). However, wake-induced interactions provide a pronounced increase in thrust-power generated by the midsection for \Aeta (compare Supplementary Figs.~S\ref{fig:pThrustEnvelopeAeta} and~S\ref{fig:pThrustEnvelopeSeta}, $0.2<s/L<0.4$).  Among all of the four swimmers compared, only \Aeta shows a distinct negative $P_{Def}$ region close to the head ($s<0.2L$), which further supports the occurrence of head-motion synchronization with flow-induced forces, when efficiency is maximized. Comparing the deformation- and thrust-power distribution for \Ady and \Sdy in Supplementary Fig.~S\ref{fig:periodPowerDistribYdisp} provides additional evidence that wake-interactions have a marked impact on swimming-energetics.

\paragraph*{\MovieOne.} 3D simulation of three nonautonomous swimmers, in which the leader swims steadily, and the two followers maintain specified relative positions such that they interact favourably with the leader's wake. The flow-structures have been visualized using isosurfaces of the Q-criterion~\cite{Hunt1988}.

\paragraph*{\MovieTwo.} 2D simulation of a pair of swimmers, in which the leader swims steadily, and the follower (\Aeta) takes autonomous decisions to interact favourably with the wake. The upper panel (labelled `$\omega$') shows the vorticity field generated by the swimmers, whereas the second panel (labelled `v') shows the lateral flow-velocity. The smart-swimmer appears to synchronize the motion of its head with the lateral flow-velocity, which allows it to increase its swimming-efficiency. The lower panels show the energetics metrics, namely, the swimming efficiency $\eta$, the thrust-power $P_{Thrust}$, the deformation-power $P_{Def}$, and the Cost of Transport (CoT).

\paragraph*{\MovieThree.} 2D simulation of a pair of swimmers, where the leader performs random actions, and the follower takes autonomous decisions to benefit from the flow-field. The smart-follower, which was trained with a steadily-swimming leader, is able to adapt to the erratic leader's behaviour without any further training. Remarkably, the follower chooses to interact deliberately with the wake in order to maximize its long-term swimming-efficiency, even though it has the option to swim clear of the unsteady flow-field.

\paragraph*{\MovieFour.} Detailed view of the flow-field around smart-swimmer \Aeta. The top panel shows the vorticity field in colour and velocity vectors as black arrows. The middle panels show the swimming-efficiency and the deformation-power. The distribution of thrust-power and deformation-power along the swimmer's left- (`lower') and right-lateral (`upper') surfaces are shown in the lower panels, and depict how these quantities depend on wake-interactions.

\paragraph*{Supplementary Movie S5.} 3D simulation of two nonautonomous swimmers, in which the leader swims steadily, and the follower maintains a specified relative position to interact favourably with the wake. The energetic-benefit for the follower is similar to that of each of the followers in \MovieOne.

\paragraph*{Supplementary Movie S6.} 3D simulation of three nonautonomous swimmers, in which the leaders use a feedback controller to maintain formation abreast of each other, and the follower holds a specified position relative to the leaders. The energetic-benefit for the follower is double that of the followers in Supplementary Movies 1 and 2, as it now interacts profitably with wake-rings generated by both the leaders.

\end{document}